\documentclass[fleqn,usenatbib]{mnras}

\usepackage{newtxtext,newtxmath}
\usepackage[T1]{fontenc}

\DeclareRobustCommand{\VAN}[3]{#2}
\let\VANthebibliography\thebibliography
\def\thebibliography{\DeclareRobustCommand{\VAN}[3]{##3}\VANthebibliography}

\usepackage{xcolor}

\usepackage{graphicx}	% Including figure files
\usepackage{amsmath}	% Advanced maths commands
\usepackage[encapsulated]{CJK}
\usepackage{ucs}
\usepackage[utf8x]{inputenc}
\newcommand{\tctext}[1]{\begin{CJK}{UTF8}{bkai}#1\ignorespacesafterend\end{CJK}}

\newcommand{\rp}{r_\mathrm{p}} % periastron distance

\title[Flybys exciting spiral arms]{Exciting spiral arms in protoplanetary discs from flybys}

\author[Smallwood et al.]{Jeremy L. Smallwood,$^{1,2}$\thanks{E-mail: jlsmallwood@asiaa.sinica.edu.tw}
Chao-Chin Yang (\tctext{楊朝欽}),$^{2,3,4}$ Zhaohuan Zhu (\tctext{朱照寰}),$^{2,3}$ \newauthor
Rebecca G. Martin,$^{2,3}$ Ruobing Dong (\tctext{董若冰}),$^5$
Nicol\'as Cuello$^{6}$ and Andrea Isella$^7$
\\
$^1$Institute of Astronomy and Astrophysics, Academia Sinica, Taipei 10617, Taiwan\\
$^2$Department of Physics and Astronomy, University of Nevada, Las Vegas, 4505 South Maryland Parkway, Las Vegas, NV 89154, USA\\
$^3$Nevada Center for Astrophysics, University of Nevada, Las Vegas, 4505 South Maryland Parkway, Las Vegas, NV 89154, USA\\
$^4$Department of Physics and Astronomy, The University of Alabama, Box~870324, Tuscaloosa, AL~35487-0324, USA\\
$^5$Department of Physics and Astronomy, University of Victoria, Victoria, BC V8P 1A1, Canada\\
$^6$Univ. Grenoble Alpes, CNRS, IPAG, 38000 Grenoble, France\\
$^7$Department of Physics and Astronomy, Rice University, 6100 Main Street, MS-108, Houston, TX 77005, USA
}
\date{Accepted XXX. Received YYY; in original form ZZZ}

\pubyear{2023}

\begin{document}
\label{firstpage}
\pagerange{\pageref{firstpage}--\pageref{lastpage}}
\maketitle

\begin{abstract}
 Spiral arms are observed in numerous protoplanetary discs.  These spiral arms can be excited by companions, either on bound or unbound orbits. We simulate a scenario where an unbound perturber, i.e. a flyby, excites spiral arms during a periastron passage. We run three-dimensional hydrodynamical simulations of a parabolic flyby encountering a gaseous protoplanetary disc. The perturber mass ranges from $10\, \rm M_J$ to $1\, \rm M_{\odot}$. The perturber excites a two-armed spiral structure, with a more prominent spiral feature for higher mass perturbers. The two arms evolve over time, eventually winding up, consistent with previous works. We  focus on analysing the pattern speed and pitch angle of  these spirals during the whole process.  The initial pattern speed of the two arms are close to the angular velocity of the perturber at periastron, and then it decreases over time. The pitch angle also decreases over time as the spiral winds up. The spirals disappear after several local orbital times. An inclined prograde orbit flyby induces similar disc substructures as a coplanar flyby. A solar-mass flyby event causes increased eccentricity growth in the protoplanetary disc, leading to an eccentric disc structure which dampens over time.  The spirals' morphology and the disc eccentricity can be used to search for potential unbound stars or planets around discs where a flyby is suspected. Future disc observations at high resolution and dedicated surveys will help to constrain the frequency of such stellar encounters in nearby star-forming regions.
\end{abstract}


%%%%%%%%%%%%%%%%%%%%%%%%%%%%%%%%%%%%%%%%%%%%%%%%%%

%%%%%%%%%%%%%%%%% BODY OF PAPER %%%%%%%%%%%%%%%%%%

\section{Introduction}
 Protoplanetary discs around young stars provide a unique laboratory for studying planet formation and  planet-disc interactions. Spiral arms, stretching from tens to hundreds of au, are observed in numerous protoplanetary discs in visible to near-infrared light.  Any perturbation in a rotating disc can excite spirals, and the spirals encode the information of the perturbation in the past \citep{Zhu2022}. Particularly, the gravitational interaction between a perturber and a disc excites strong spirals, and thus spirals are normally regarded as a signpost for undetected planets \cite[e.g.,][]{Grady1999,Muto2012,Grady2013,Wagner2015,Monnier2019,Garufi2020,MuroArena2020}. On the other hand, spirals can also be excited through flyby interactions. 
A flyby is defined as an object undergoing a single periastron passage of a stellar companion within $100-1000\, \rm au$. Stars are formed in relatively dense clusters, and therefore the interaction between stellar counterparts is enhanced \citep{Hillenbrand1997,Carpenter2000,Lada2003,Porras2003}.  From the work of \cite{Pfalzner2013} and \cite{Winter2018a}, the probability of stellar flybys are about $30\%$ for solar-type stars during the first million years of stellar evolution, using a background stellar density that is larger than in Taurus.  Considering that the spirals excited by an eccentric perturber are  dramatically different from those excited by a circular perturber \citep{Zhu2022}, studying the spirals excited by an unbound flyby perturber is desired to understand the origin of the observed spirals and how planetary systems are formed.

Flyby encounters have been both simulated numerically and observed in protoplanetary discs. \cite{Cuello2019} ran extensive 3-dimensional smoothed particle hydrodynamic simulations of protoplanetary discs (consisting of both gas and dust) being perturbed by a stellar flyby on a parabolic orbit. The authors explored a variety of different orbital inclinations, periastron distances, and mass ratios of the flyby event. They find that spiral structures are produced in both gas and dust, which have a lifetime of the order of thousands of years.  The perturber truncates  and warps the primary disc for a range of perturber inclinations and periastron distances \citep{Clarke1993,Ostriker1994,Terquem1996,Bhandare2016,Xiang-Gruess2016}, leaving the dust disc more compact than the gas disc. Furthermore, there are multiple observational signatures that are present during a recent flyby encounter.  Aside from spiral formation, long bridges of material are linked from the primary disc to the intruding flyby \citep{Cuello2020}.  \cite{Clarke1993} demonstrated that a prograde, coplanar parabolic flyby encounter stripped material off the protoplanetary disc, and the perturber captured a portion of the stripped material. Warps and misalignments are common in the primary disc and are observable in moment one maps. Broken protoplanetary discs can have large mutual misalignments between the inner and outer gas rings that may be generated by a flyby scenario \citep{Nealon2020}. There have been several cases of flyby candidates in RW Aur \citep{Dai2015}, HV Tau and DO Tau \citep{Winter2018b}, AS 205 \citep{Kurtovic2018}, UX Tauri \citep{Menard2020}, Z CMa \citep{Dong2022}, and FU~Ori \citep{Borchert2022}. For a recent review on flyby's shaping protoplanetary discs, see \cite{Cuello2023}.

Stellar flybys may also be responsible for shaping wide planetary systems \citep{Bailey2019,cai2019,Batygin2020,Li2020,Wang2020a,Wang2020b}, including our solar system \citep{Kobyashi2005,Pfalzner2018}. Stellar flybys can strongly perturb protoplanetary discs during their periastron passage, which can impact the total size and frequency of planetary systems \citep{Scally2001,Kenyon2002,Olczak2006,Steinhausen2014,Rosotti2014,PortegiesZwart2016,Vincke2016,ConchaRamirez2019,Jumenez2020}.  Most of these stellar flyby encounters are especially weak due to the fast relative velocity \citep{VerasMoeckel2012}. Still, low relative velocity encounters occurring within dense environments have a more dominant gravitational interaction \citep{li2015}. Flyby encounters are also hypothesized to excite the migration of outer giant planets, forming hot Jupiters \citep{Rodet2021}. The origin of the highly inclined Trans-Neptunian objects may also be associated with stellar flybys \citep{Moore2020}.

  Free-floating (rogue) planets may also perturb discs by flyby interactions \citep{Gladman2006}. A planet can become ejected from its parent system by planet-planet scattering \citep{Rasio1996,Lin1997,Chatterjee2008}. $75$ per cent of exoplanetary system's with giant planets are expected to undergo planet-planet scattering \cite[e.g.,][]{Raymond2022}. It is found to be easier to eject rocky planets rather than giant planets \citep{Veras2005,Matsumura2013,Carrera2016}, however ejection of giant planets can still occur. Rogue planets are more common to form around high mass stars \citep{Ma2016,Barclay2017}. Recently, \cite{Mroz2020} found a free-floating planet candidate from microlensing events. Other mechanism for producing rogue planets include: interactions in multi-star systems \citep{Kaib2013,Spurzem2009}, stellar flybys \citep{Malmberg2011}, or post-main-sequence evolution \citep{Veras2011}. 

\begin{table*}
	\centering
	\caption{The setup of the SPH simulations. The simulation ID is given in the first column. The remaining columns lists the tilt of the incoming perturber orbit, $i_2$,
	%\nc{this should be called $r_{\rm p}$ perhaps to avoid confusion, see comment Eq. 6 for instance. $r_2$ is the distance between the stars.}
	the initial mass of the flyby intruder, $M_2$, the mass ratio of the intruder to the central star, $M_2/M_1$, the number of particles, rotation axis, and whether or not spirals are generated. }
%	\label{tab:example_table}
	\begin{tabular}{ccccccc} % four columns, alignment for each
		\hline
		Simulation ID & $i_{2}/^\circ$ & $M_2/M_{\rm J}$ & $M_2/M_1$ & $\#$ of particles & Rotation axis& Spirals?   \\
		\hline
		\hline
        %M5 & $0$ & $200$ & $5$ & $0.005$ & $10^6$  & &  Yes\\
        M10 & $0$  & $10$ & $0.01$ & $10^6$  & -- & Yes\\
        %M20 & $0$ & $200$ & $20$ &  $0.02$ &  $10^6$ & & Yes\\
        M100 & $0$ & $100$ & $0.1$  & $10^6$  & -- & Yes\\
        M100h & $0$  & $100$ & $0.1$ & $10^7$ &--  &Yes \\
        M1000 & $0$  & $1000$   & $1$ &  $10^6$ &-- & Yes\\
 %       run7 & $0$ & $120$ & $100$ & $0.1$ & $10^6$  & & Yes \\
 %       run8 & $0$ & $180$ & $100$ & $0.1$ & $10^6$ & & Yes \\
 %       M100r & $180$ & $200$ & $100$ & $0.1$ & $10^6$ & & No\\
        M100iy & $45$  & $100$ & $0.1$ & $10^6$ & $y$--axis & Yes\\
        M100ix & $45$  & $100$ & $0.1$ & $10^6$ & $x$--axis & Yes\\
        \hline
	\end{tabular}
    \label{table::setup}
\end{table*}

Protoplanetary discs around single stars are observed to have a lifetime of around $1-10\, \rm Myr$ \citep{Haisch2001,Hernandez2007,Hernandez2008,Mamajek2009,Ribas2015}. These lifetimes are long enough for discs to be perturbed by stellar flyby encounters \citep{Cuello2019,Jimenez-Torres2020}.  As the perturber approaches periastron passage,  tidal effects  prompt the formation of spirals \citep{Ostriker1994,Pfalzner2003,Shen2010,Thies2010}. External  companions will excite spiral density waves at Lindblad resonances,  which has been investigated analytically \citep{Goldreich1978,Goldreich1979,Lin1993,Lubow1998,Ogilvie2002} and numerically \citep{Kley1999,Dong2011,Zhu2015b,Dong2017,Hord2017,Bae2018,Zhang2020,Ziampras2020,Muley2021}. Spiral are launched at these resonances because the local orbital frequency is a multiple of the Doppler-shifted forcing frequency of the companion. If the companion is an external star, it exerts a strong tidal force where its Roche lobe can reach beyond the location of most of these resonances.  The superpostion of density waves excited at Lindblad resonances can constructively interfere, leading to the production of multiple spiral arms \citep{Bae2018,Miranda2019}. In the case of a companion on an unbound orbit, density waves are not repeatedly excited, leading to the arms winding up overtime \cite[e.g.,][]{Pringle2019,Semczuk2017,Kumar2022}.

Spiral density waves can be described by two main parameters, their pattern speed and their pitch angle. The pattern speed is how quickly a spiral arm appears rotating azimuthally, and the pitch angle is the angle between the tangent to a spiral arm and a perfect circle, which measures how tightly the spiral arms are wound. The pattern speed and pitch angle of spiral arms in protoplanetary discs can be measured from observations \citep{kraus2017,Dong2018b,Huang2018,Casassus2021}.

 This work further examines the formation and evolution of spiral arms in protoplanetary discs that are produced by flyby encounters. We systemically measure the pattern speed and pitch angle of the excited spiral arms in discs around a solar mass star for different perturber masses in the range of $M_2 = 10-1000\, \rm M_{\rm J}$, where $M_{\rm J}$ is Jupiter mass. The evolution of the pattern speed and pitch angle can be compared to observations in order to infer the presence of unbound perturbers.  The origin of spirals has direct implications for both planet formation and disc evolution \citep{Dong2018a,Brittain2020}. The layout of the paper is as follows. In Section~\ref{sec::1 Methods}, we explain the setup of the hydrodynamical flyby simulations and showcase the results in Section~\ref{sec::Results}. In Section~\ref{sec::Discussion}, we discuss how our results are related to the observed properties of the spiral arms seen in various systems. Lastly, we draw our conclusions in Section~\ref{sec::Conclusions}.

\section{Methods}
\label{sec::1 Methods}
To model a flyby of a massive object perturbing a protoplanetary disc, we use the 3-dimensional smoothed particle hydrodynamical (SPH) code {\sc phantom} \citep{Price2018,Price2010}. The code can model a wide variety of geometries of parabolic orbits. On top of this, the system's angular momentum is conserved with  the same order of accuracy as that of the time-stepping scheme. In {\sc phantom}, linear and angular momentum are both conserved to round-off error, typically $\sim 10^{-16}$ in double precision. Recently, {\sc phantom} has been well tested for parabolic flyby encounters interacting with gaseous and dusty discs \citep{Cuello2019,Cuello2020,Nealon2020,Menard2020}.  The reference frame within our simulations is centered on the center-of-mass of the system. A summary of the simulations is given in Table~\ref{table::setup}.  The simulation ID is given in the first column, where the "M" corresponds to the mass of the perturber and the numbers following give the mass value in units of Jupiter mass ($M_{\rm J}$). For example, M100 is a $100\, \rm M_J$ perturber. We also consider a higher resolution case, M100h, and inclined encounters, M100iy and M100ix. 

\subsection{Circumprimary disc setup}
We set up a coplanar gaseous protoplanetary disc around a generic solar-type star.  The hydrodynamical disc is in the bending waves regime where the disc aspect ratio $H/r$ is larger than the viscosity coefficient $\alpha$ \citep{Shakura1973}. Within this regime, the warp induced by the interaction of the flyby propagates as a pressure wave with speed $c_s/2$ \citep{Papaloizou1983,Papaloizou1995,Lubow2000}, where $c_{\rm s}$ is the sound speed. We model  an initially flat disc with $10^6$ Lagrangian particles with a total disc mass of $0.001\, \rm M_{\odot}$ 
%\nc{Is this 1\% or 0.001?}.
We do, however, include one higher resolution simulation with $10^7$ particles (see Table~\ref{table::setup}). The low disc mass has a negligible dynamical effect on the orbit of the flyby. The mass of the primary star is $M_1 = 1\, \rm M_{\odot}$. The inner disc radius is initially set to $r_{\rm in} = 10\, \rm au$ and the outer radius is $r_{\rm out} = 100\, \rm au$.  The disc lies initially in the $x$--$y$ plane. The accretion radius for the central star is set to $10\, \rm au$. The accretion radius is a hard boundary. Any Lagrangian particle that passes across the boundary is considered accreted, and the particle's mass and angular momentum are deposited onto the star. Our primary sink accretion radius is comparable to the initial inner edge of the disc to reduce the computational time by neglecting to resolve close-in particle orbits. The surface density profile is initially a power-law distribution given by
 \begin{equation}
     \Sigma(r) = \Sigma_0 \bigg( \frac{r}{r_{\rm in}} \bigg)^{-p} 
    \bigg(1-\sqrt{\frac{r_{\rm in}}{r}}\bigg),
     \label{eq::sigma}
 \end{equation}
where $\Sigma_0 = 7.00\, \rm g/cm^2$ is the density normalization and $p$ is the power law index. We set $p=+3/2$ 
%\nc{just note that most of the previous work on flybys with phantom was done using p=1 if I am not mistaken. This fact could be useful for the discussion.}
and the density normalization is defined by the total disc mass. We use a locally isothermal equation-of-state with a disc thickness that is scaled with radius as
\begin{equation}
    H = \frac{c_{\rm s}}{\Omega} \propto r^{3/2-q}, 
 \end{equation}
where $\Omega = \sqrt{GM_1/r^3}$ and $c_{\rm s}$ is proportional to $r^{-q}$.  We use $q = +1/2$ so that the initial disc aspect ratio  is constant at $H/r = 0.05$.

 The \cite{Shakura1973} viscosity, $\alpha_{\rm SS}$, prescription is given by 
\begin{equation}
    \nu = \alpha_{\rm SS} c_{\rm s} H,
\end{equation}
where $\nu$ is the kinematic viscosity. To calculate $\alpha_{\rm SS}$, we use the prescription in \cite{Lodato2010} given as
\begin{equation}
\alpha_{\rm SS} \approx \frac{\alpha_{\rm AV}}{10}\frac{\langle h \rangle}{H},
\end{equation}
where $\langle h \rangle$ is the mean smoothing length on particles in a cylindrical ring at a given radius \citep{Price2018}. We take the \cite{Shakura1973} viscosity parameter to be $\alpha_{\rm SS} = 0.005$, which sets the artificial viscosity to $\alpha_{\rm AV} = 0.1260$ ($0.2713$ for the high-resolution simulation).  Note that a value of $\alpha_{\rm AV} = 0.1$ represents the lower limit, below which a physical viscosity is not well resolved in SPH, meaning that viscosities smaller than this lower limit produce disc spreading independent of the value of $\alpha_{\rm AV}$ \cite[see][for details]{Meru2012}.  In hydrodynamics, we also include a parameter, $\beta_{\rm AV}$, which provides a non-linear term that was originally introduced to prevent particle penetration in high Mach number shocks \cite[e.g.,][]{Monaghan1989}. In general, $\beta_{\rm AV} = 2.0$.  In {\sc phantom}, the smoothing length is specified as a function of the particle number density $n$ given as
\begin{equation}
    h = h_{\rm fact}n^{-1/3},
\end{equation}
where $h_{\rm fact}$ is the proportionality factor specifying the smoothing length in terms of the mean local particle spacing. The disc is resolved with a shell-averaged smoothing length per scale height of $\langle h \rangle/H \approx 0.39$ and $\langle h \rangle/H \approx 0.18$ for our high-resolution simulation. Lastly, we ignore the effect of disc self gravity.

\subsection{Flyby setup}
We consider an unbound perturber that gravitationally influences a gaseous protoplanetary disc around the primary star. We simulate only parabolic encounters ($e = 1$), which induce the greatest star-to-disc angular momentum transfer and therefore engender the most prominent spiral features in the disc \citep{Vincke2016,Winter2018b,Cuello2019}. Perturbers on hyperbolic trajectories ($e>1$) would have a higher angular velocity at the closest approach, leading to a less significant impact on the generation of structures within the disc \cite[e.g.,][]{Winter2018b}. Hyperbolic orbit flybys are also less efficient in capturing material compared to parabolic encounters \cite[e.g.,][]{Larwood1997,Pfalzner2005}.

  We model flybys on prograde coplanar and inclined trajectories. The center of mass of the system is set as the center of the $x$--$y$ plane. We measure the tilt of the flyby orbit with respect to the $z$--axis. The distance between the perturber and the central star is defined as $r_2$. A coplanar perturber initially lies in the $x$--$y$ plane (perturber's angular momentum vector is parallel to the $z$--axis), arrives initially from the negative $y$ direction, and leaves towards the negative $y$ direction. We also consider two prograde inclined trajectories that are tilted by $45^\circ$. The inclined models are rotated clock-wise with respect to the $y$--axis (model M100iy) and $x$--axis (model M100ix), respectively. For the coplanar and y-axis rotation cases, the 3D position of the pericentre does not change (similar to the simulations from \cite{Cuello2019}) but that for the x-axis rotation the pericentre position is sent out of the disc plane. This may affect the excitation mechanism and the efficiency in terms of angular momentum transfer.

 The probability of an encounter with a periastron distance $100-1000\, \rm au$ is roughly $30\%$ but rapidly decreases for increasing stellar cluster age for leaky cluster environments \cite[e.g.,][]{Pfalzner2013}. Therefore,  we select the minimum separation between the primary and the perturber (periastron distance) to be $r_{\rm p} = 200\, \rm au$, and hence the distance between the perturber and the center of mass at the closest encounter varies with perturber mass. The periastron of the coplanar and  $y$--axis rotated simulations occurs at $x=0$, $y>0$, and $z=0$. For the $x$--axis rotated simulation, the periastron occurs at $x=0$, $y>0$, and $z<0$.  We consider perturber masses of $10\, \rm M_{J}$, $100\, \rm M_{J}$, and $1\, \rm M_\odot$. The accretion radius of the perturber is set to $1\, \rm au$.

\section{Results}
\label{sec::Results}
We analyze the primary protoplanetary disc both radially and azimuthally  using spherical shells. For our radial analysis, we divide the disc into $300$ linear bins in spherical radius, $r$, which range from $10\, \rm au$ (the initial inner disc edge) to $100$ au (the initial outer disc edge). For our azimuthal analysis, we use $250$ bins that range from $0\leq \theta \leq 2\pi$ at a given radius $r$, where $\theta$ is the azimuthal angle. Within each bin, radial or azimuthal, we calculate the mean properties of the particles, such as the surface density, inclination, longitude of ascending node, and eccentricity.

\begin{figure} 
\centering
\includegraphics[width=\columnwidth]{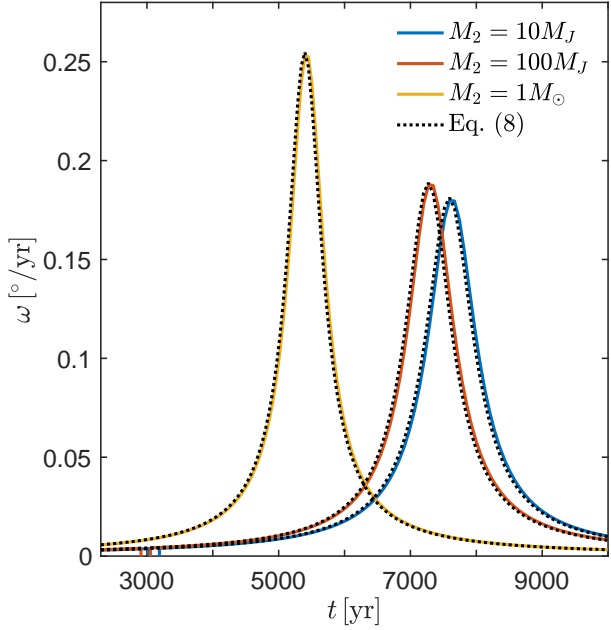}
\centering
\caption{The angular velocity of the flyby perturber with different masses from the hydrodynamical simulations. We consider perturber masses of $10\, \rm M_J$ (blue, M10),  $100\, \rm M_J$ (red, M100), and $1\, \rm M_{\odot}$ (yellow, M1000). The peak of the angular velocity (solid lines) occurs when the perturber is at periastron. The analytically derived angular velocities for each perturber mass from Eq.~(\ref{eq::omega}) are given by the dotted curves.  The effect of perturber mass can change the angular velocity and the time of periastron passage.  }
\label{fig::flyby_speed}
\end{figure}

\begin{figure} 
\centering
\includegraphics[width=\columnwidth]{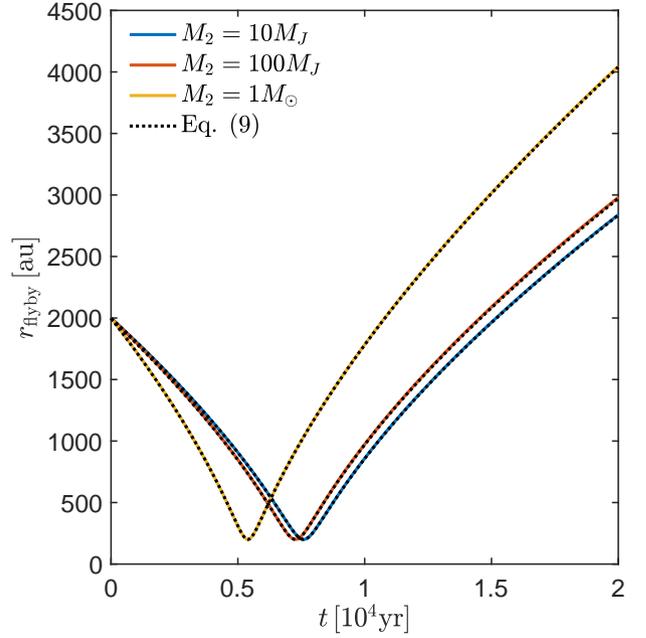}
\centering
\caption{The separation of the unbound perturber to the primary for different perturber masses. We consider perturber masses of $10\, \rm M_J$ (blue, M10),  $100\, \rm M_J$ (red, M100),  and $1\, \rm M_{\odot}$ (yellow, M1000). The trough of the separation curves correspond to the time of periastron passage, which occurs at a distance of $200\, \rm au$.  The analytically derived separation for each perturber mass from Eq.~(\ref{eq::rp}) are given by the dotted curves. The more massive perturber arrives at periastron earlier than the less massive perturbers.  }
\label{fig::flyby_dist}
\end{figure}

\subsection{Flyby orbit}
We now examine the angular velocity and distance of the  flyby encounters during the simulations.  The parabolic orbit in our model is described by
\begin{equation}
    r_2 = \frac{2\rp}{1 + \sin\theta},
\end{equation}
where $\rp$ is the periastron distance, which occurs at $\theta = +\pi / 2$.
The angular speed as a function of $r_2$ is then
\begin{align}
    \omega(r_2) &=
    \sqrt{\frac{2G(M_1 + M_2)}{\rp^3}}
    \left(\frac{\rp}{r_2}\right)^2\\&=
    \left(0\fdg18\,\textrm{yr}^{-1}\right)
    \left(\frac{M_1 + M_2}{M_{\sun}}\right)^{1/2}
    \left(\frac{r_\mathrm{p}}{200\,\textrm{au}}\right)^{-3/2}
    \left(\frac{\rp}{r_2}\right)^2,
    \label{eq::omega}
\end{align}
where $G$ is the gravitational constant, $M_1$ is the mass of the primary star, and $M_2$ is the mass of the perturber. The relationship between the  radial distance, $r_2$, and time $t$ is given by
\begin{equation}
    \left(\frac{r_2}{r_{\rm p}} + 2 \right) \sqrt{\frac{r_2}{r_{\rm p}} - 1} = \frac{3|t - t_\mathrm{p}|}{2} \sqrt{\frac{2G (M_1 + M_2)}{r_{\rm p}^3}},
    \label{eq::rp}
\end{equation}
where $r_2 = r_\mathrm{p}$ when $t = t_\mathrm{p}$.
Figure~\ref{fig::flyby_speed} shows the angular velocity of the perturber for masses $M_2 =$ $10\, \rm M_J$, $100\, \rm M_J$, and $1\, \rm M_\odot$. The peak of the angular velocity occurs when the perturber is at periastron. The more massive the perturber, the higher the angular velocity at periastron, and the time of periastron passage occurs sooner. The derived angular speed for each parabolic companion from Eq.~(\ref{eq::omega}) is shown by the black-dotted curves, which is consistent with the simulation results.  Figure~\ref{fig::flyby_dist} shows the separation of the unbound perturber to the primary star. The troughs in each model represent the time at which the flyby reaches periastron, which is at a distance of $200\, \rm au$  by construction.  The derived radial separation for each parabolic companion from Eq.~(\ref{eq::rp}) is shown by the black-dotted curves, which is consistent with the simulation results.

\begin{figure}
\centering
\includegraphics[width=0.97\columnwidth]{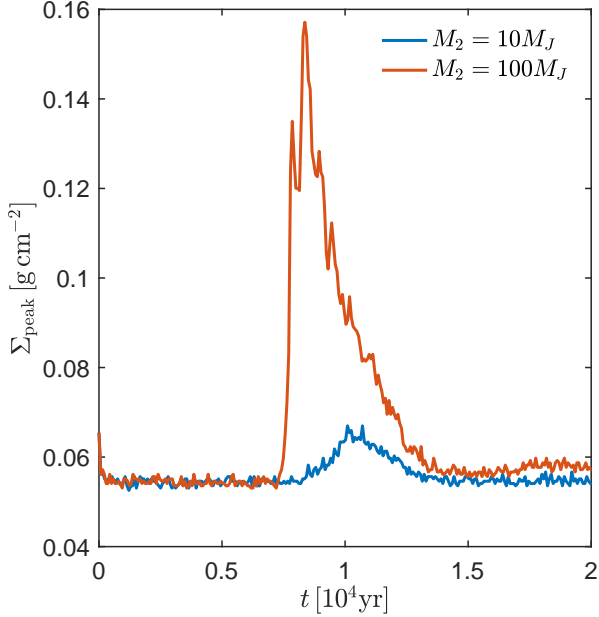}
\centering
\caption{The peak disc surface density, $\Sigma_{\rm peak}$,  as a function of time.  The peak density is found by taking the maximum surface density over the disc azimuthal angle ($0 < \theta < 2\pi$) at $r = 65\, \rm au$.  We consider perturber masses of $10\, \rm M_J$ (blue, M10), and  $100\, \rm M_J$ (red, M100).  A more massive perturber excites denser spiral arms.  %\RGM{Is this a maximum over varying azimuthal angle?}
}
\label{fig::sigma_peak}
\end{figure}

\begin{figure*} 
\centering
\includegraphics[width=1.9\columnwidth]{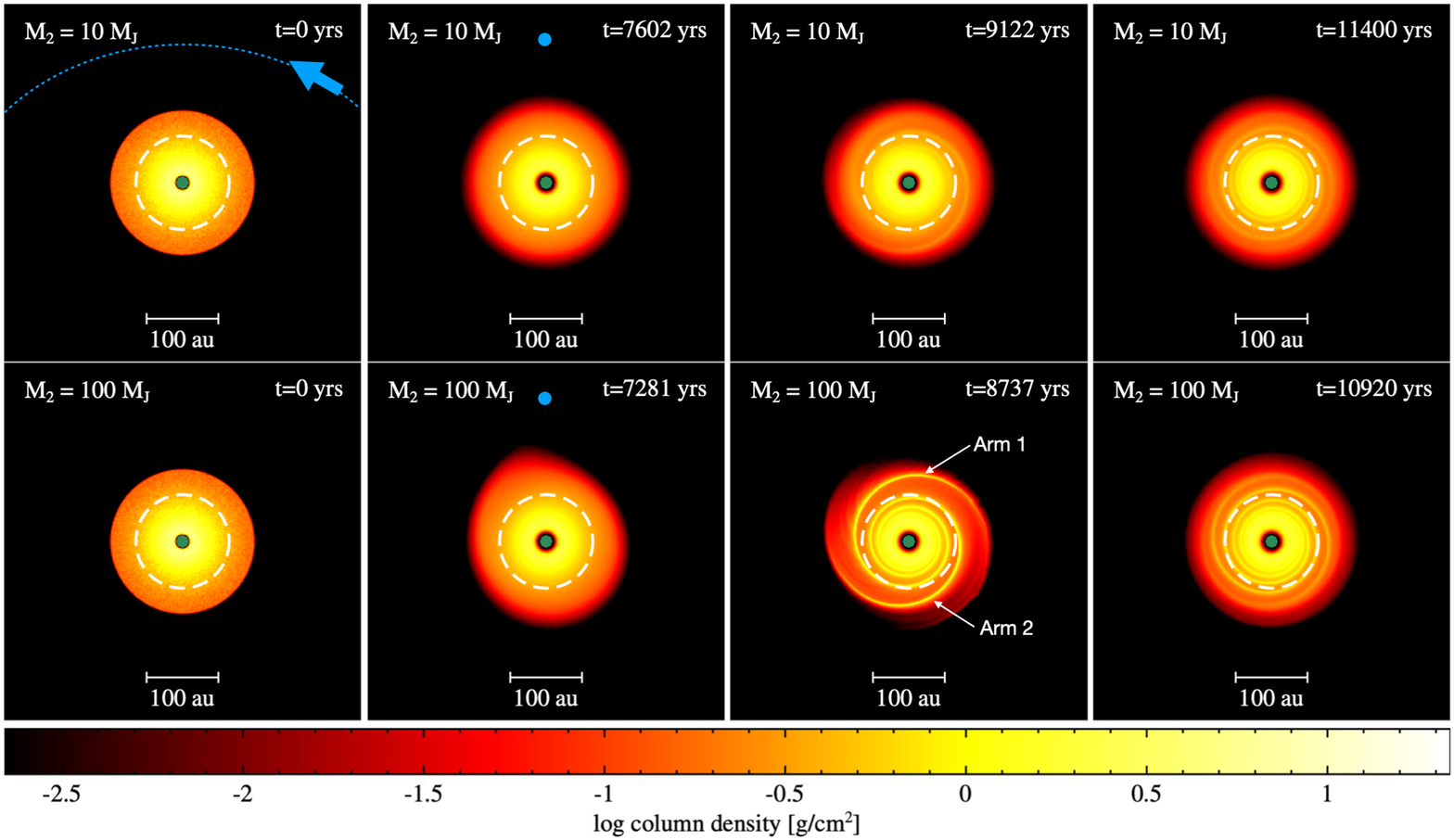}\centering
\centering
\caption{Evolution of the spiral structure for two different perturber masses, $10\, \rm M_J$ (M10, first row), $100\, \rm M_{J}$ (M100, second row). We show the face-on disc structure viewed in the $x$--$y$ plane. For visualisation purposes only, we centre the coordinates and velocities on the primary star (green dot). We show the disc structure at four different times, with the first column being the initial structure of the disc, with the parabolic trajectory of the perturber given by the blue dotted curve. and the second column being the time of closest approach of the perturber (blue dot). In the third and fourth columns, the perturber is outside the field of view.   A more massive intruding flyby produces a more prominent two-armed spiral structure that evolves over time. A depiction of the two-armed structure is shown in the bottom row, third panel, where arm 1 is initially excited closest to the perturber and arm 2 is initially excited furthest from the perturber. Throughout this work,  we probe the disc properties at $65\, \rm au$, shown by the dashed-white circle. }
\label{fig::splash}
\end{figure*}

\subsection{Coplanar non-equal mass flybys} \label{SS:coplaner_lowmass}

In this subsection, we first analyse the  simulations with a non-equal mass perturber, models M10 and M100 from Table~\ref{table::setup}. In Fig.~\ref{fig::sigma_peak}, we plot the peak surface density, $\Sigma_{\rm peak}$,  at a radius $r = 65\, \rm au$ as a function of time for $M_2 = 10,\,100\, \rm M_J$.  The peak surface density is defined as the maximum surface density value at each time. At $r = 65\, \rm au$, we have relatively stronger density contrast near the arms. The maximum point of the curves corresponds to the strength of the spiral arms that are produced. 
%The more massive perturber excites a peak density that is roughly a factor of two greater than the lower mass perturber. 
 The density enhancement for the $10\, \rm M_J$ model is a factor of $\sim 1.2$, while it is a factor of $\sim 3$ for the $100\, \rm M_J$ model. Next, we visualize the spiral structure that is excited within the disc. Figure~\ref{fig::splash} shows the disc evolution at four different times for a $10\, \rm M_J$ (first row, M10), and $100\, \rm M_J$ (second row, M100) flyby encounter.  For visualization purposes only, the coordinates and velocities are centered on the primary star, marked with a green dot at $x = 0$ and $y=0$. The first column shows the discs at $t=0\, \rm yr$, with a trace of the perturber's parabolic orbit given in the top panel. The second column shows the flyby at periastron (or closest approach), seen by the blue dot to the north. At this time, the perturbers have not excited any substructures in each case. The third and fourth columns show times shortly after periastron passage, where the perturbers  excite a two-armed spiral structure in the disc. A lower mass perturber excites weaker spiral arms than a higher-mass perturber. The arms begin to wind up on themselves  when the perturber moves away from the primary star. The spiral arms generated from a more massive companion is more prominent -- leading to a higher surface density concentration within the arms, as shown in Fig.~\ref{fig::sigma_peak}. When a two-armed spiral is excited, the arm that is excited initially closest to the perturber is denoted as "arm 1", and the arm excited initially furthest from the perturber is "arm 2". A depiction of arms 1 and 2 are given the bottom row, third panel in Fig~\ref{fig::splash}.

\begin{figure*} 
\centering
\includegraphics[width=2\columnwidth]{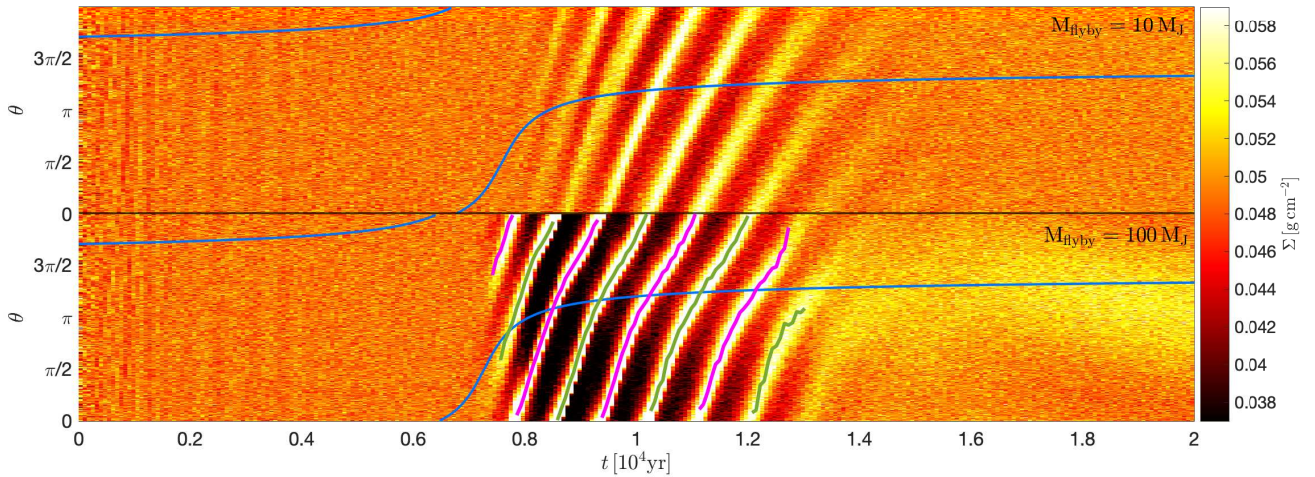}
\centering
 \caption{An azimuthal cut of the disc surface density as a function of time in years. The colour maps denote the disc surface density. We analyze the azimuthal angle $\theta$ at $r = 65\, \rm au$.  The azimuthal angle for the flyby orbit is given by the blue line. The time of closest approach occurs at $\theta = +\pi/2$. The top panel denotes a $10\, \rm M_J$ mass perturber, and bottom panel is a $100\, \rm M_J$ mass perturber. The more massive perturber excites more prominent spiral arms.  The magenta and green lines in the bottom panel trace arm 1 and arm 2, respectively, by the procedure described in Section~\ref{SS:coplaner_lowmass}. The same tracing procedure is applied to all other simulations. }
\label{fig::space_time}
\end{figure*}

\begin{figure}
\centering
\includegraphics[width=0.99\columnwidth]{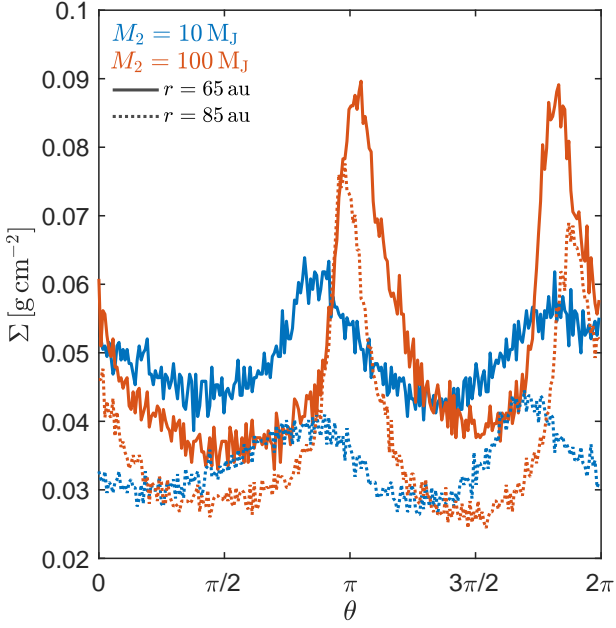}
\centering
\caption{ An azimuthal cut showing the arm contrast versus azimuthal angle at $t = 10^4\, \rm yr$ for a $10\, \rm M_J$ perturber (blue) and $100\, \rm M_J$ perturber (red). We show the arm contrast at two different radii, $65\, \rm au$ (solid lines) and $85\, \rm au$ (dotted lines). The two peaks in the surface density correspond to arms 1 and 2 from left to right.}
\label{fig::azimuthal_cut}
\end{figure}

\begin{figure}
\centering
\includegraphics[width=0.99\columnwidth]{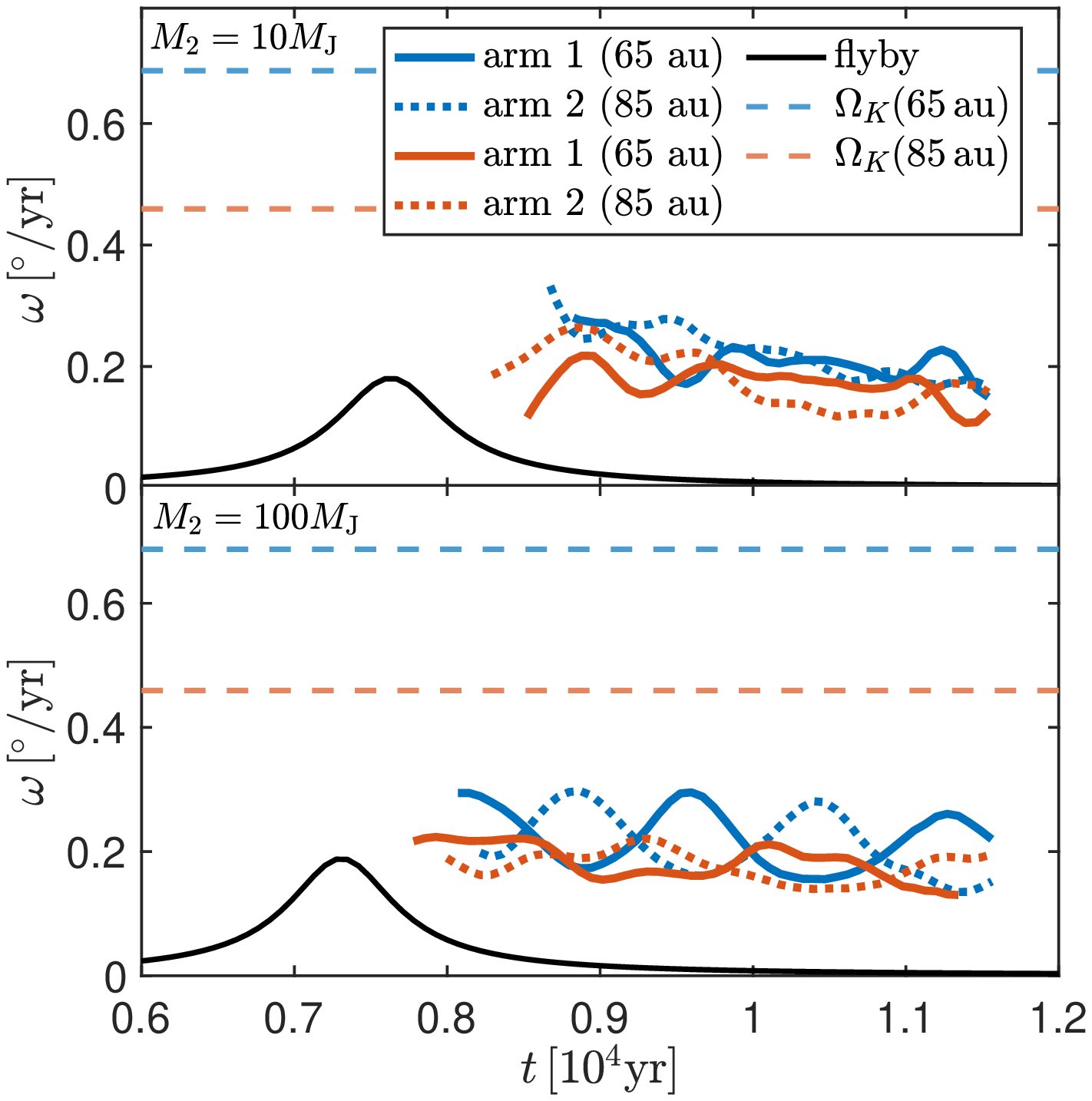}
\centering
\caption{The pattern speed, $\omega$, of the two-armed spiral at two different radii, $r = 65\, \rm au$ (blue) and $r = 85\, \rm au$ (red) for $M_2 = 10\, \rm M_{\rm J}$ (top panel, M10) and  $M_2 = 100\, \rm M_{\rm J}$ (bottom panel, M100).  We smooth the data using a Gaussian filter over a ten-element sliding window. The pattern speed for arm 1 is given by the solid curves, and arm 2 by the dotted curves. The angular velocity of the perturber is shown by the black curve. The peak corresponds to the time of closest approach of the flyby. The Keplerian angular velocity, $\Omega_{\rm K}$, at $r = 65\, \rm au$ and $r = 85\, \rm au$ are given by the blue and red horizontal dashed lines, respectively.  The pattern speeds of the spiral arms are initially closer to the angular velocity of the flyby than the Keplerian angular velocity. 
%\rdnote{For a moment i couldn't find the two horizontal Keplerian lines as they are close to the legend. Not sure if this is a problem for others. One potential treatment is to move the legend out of the figure box. Also, the figure is clear once one stares at it for a minute, though it's a bit busy. It could be potentially split into two if you feel the same (e.g., separate M100 and M10).}
}
\label{fig::pattern_speed}
\end{figure}

% \begin{figure}
% \centering
% \includegraphics[width=\columnwidth]{spiral_trace.eps}
% \centering
% \caption{The tracing procedure for identifying arm 1  (blue) and arm 2 (green) at $r = 65\, \rm au$ for the simulation with a $100\, \rm M_J$ flyby.  The tracing procedure is robust, and can be applied to all other simulations. }
% \label{fig::spiral_trace}
% \end{figure}

 We now examine the azimuthal  variation of the surface density  at a given radius for each coplanar non-equal mass perturber simulation. Figure~\ref{fig::space_time} shows the azimuthal cut of the disc surface density at a radius of $r = 65\, \rm au$. The top panel shows the disc structure under the influence of a $10\, \rm M_J$ (M10) perturber, and the bottom panel for a $100\, \rm M_J$ (M100) perturber. The angular position of the flyby is given by the solid blue curve. The time of periastron passage occurs at $\theta = +\pi/2$ in each simulation. Before the time of the closest approach of the perturber, the protoplanetary disc  appears quiescent, with no excited substructures. Soon after periastron passage, the perturber excites a two-armed spiral within the disc, seen by the  azimuthal striations in the surface density. From our comparison of different unbound-perturber masses, we find that the more massive perturber excites more prominent arms versus a lower-mass perturber. The density enhancement between the two cases is about a factor of two, which we note can be distinguishable by observations (see Section~\ref{sec::Discussion}). On average, the spiral arms in each simulation last for approximately $5000\, \rm yr$  (equivalent to $\sim2$ Keplerian orbits at $r_{\rm p}$) after the closest approach of the flyby,  as shown in Fig.~\ref{fig::space_time}.  When the spiral arms fully wind up, the resulting disc structure restores to the initial disc structure.

  In Fig.~\ref{fig::azimuthal_cut}, we display the spiral arm contrast versus azimuthal angle for a $10\, \rm M_J$ perturber (blue lines) and $100\, \rm M_J$ perturber (red lines). The solid lines denote the spiral arm at radius $r = 65\, \rm au$, and the dotted lines at $r = 85\, \rm au$. The two peaks in the surface density represent the individual arms. For example, for $M_2 = 100\, \rm M_J$, arm 1 is located just beyond $\theta \sim \pi$, and arm 2 is located just below $\theta \sim 2\pi$.  The density contrast between the arms is identical for each perturber mass at each radius.

 In order to map out arm 1 and arm 2, we apply a tracing technique shown in the bottom panel of  Fig.~\ref{fig::space_time}. At each time step, we track highest density located within the spirals. We continue tracing the arms until both arms cannot be fully traced. We then store the azimuthal angle and time that comprise the spiral arms. The tracing procedure is robust, and can be applied to all other simulations. We begin the analysis of the pattern speed and pitch angle at $t = 8000\, \rm yr$ and track the properties of the spiral arms until $t = 11500\, \rm yr$. This selected time range ensures that the spiral arms are better defined, resulting in a high-precision analysis.

 The pattern speed of an arm is the derivative of its aziuthmal position at a given radius with respect to time.  We compute the pattern speeds of the spiral arms by measuring the slope of the curves from our spiral arm tracings. Figure~\ref{fig::pattern_speed} shows the pattern speed of arm 1 (solid curve) and arm 2 (dotted curve) as a function of time  for each perturber mass. We calculate the pattern speed at two different radii, $65\, \rm au$, and  $85\, \rm au$. We smooth the data using a Gaussian filter. The pattern speed of the spiral arms are higher closer to the central star. The angular velocity of the perturbers are shown by the black and gray curves and the Keplerian angular velocity at radii $r = 65\, \rm au$ and $85\, \rm au$  are shown by the horizontal dashed lines.  The initial pattern speeds of the arms are closely set by the angular velocity of the perturber at periastron. %However, the presence of the Keplerian velocity causes the initial pattern speed to be higher than the perturber's peak angular velocity. 
The pattern speed of the spiral arms decreases with time and the arms ultimately dissipate $\sim 5$ thousand years after the closest approach of the flyby.  For a given radius,  the pattern speed of arm 1 and arm 2 oscillate  out of phase with respect to one another. This phenomenon can be seen clearly in the blue curves from Figure~\ref{fig::pattern_speed}, where the trough of one arm  occurs at the peak of the other arm. The oscillation amplitude of the arms induced by a lower mass perturber is lower than the oscillations driven by a higher mass perturber.  We tested this phenomenon with a bound perturber, finding the pattern speed of the arms does not display such oscillations but instead has a constant pattern speed at a given radius, as expected, thus ruling out numerical causes of the phenomenon. We defer further exploration of this phenomenon to a future work.
%(Also, at this point readers may likely wonder ``Hmmm this is strange. Do you have an explanation?'' To save them from keeping this question as they move forward, you could potentially add something like ''It is unclear what causes the fluctuation'' or ''We defer further exploration of this phenomenon to a future work''.)}.}

%\RGM{5000 years doesn't seem long lived to me. Especially since you have picked parameters to get the maximum impact, e.g. the trajectory. Maybe change the title?}

\begin{figure} 
\centering
\includegraphics[width=\columnwidth]{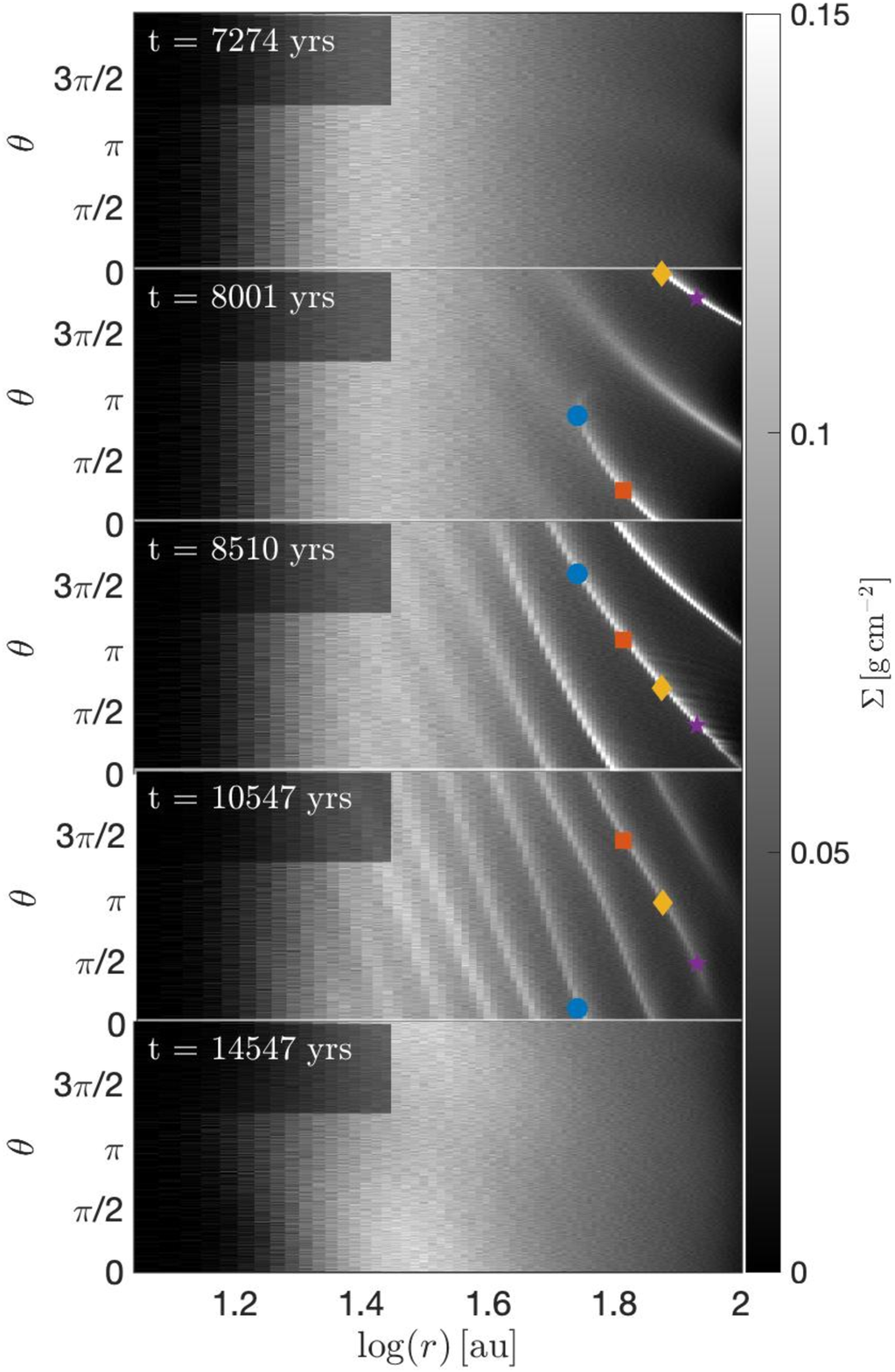}
\caption{ Surface density in an ($r$,$\theta$) grid. The maps show the surface density in units of $\rm g\, cm^{-2}$. We show the surface density at  five different times, with the upper panel being the time when the perturber is at closest approach. We mark the peak surface density along arm 1 at four radii, $55\, \rm au$ (blue circle), $65\, \rm au$ (red square), $75\, \rm au$ (yellow diamond), and $85\, \rm au$ (purple pentagram). These spiral arms eventually wind up,  as shown in the bottom panel.}
\label{fig::pitch_spirals}
\end{figure}

\begin{figure} 
\centering
\includegraphics[width=\columnwidth]{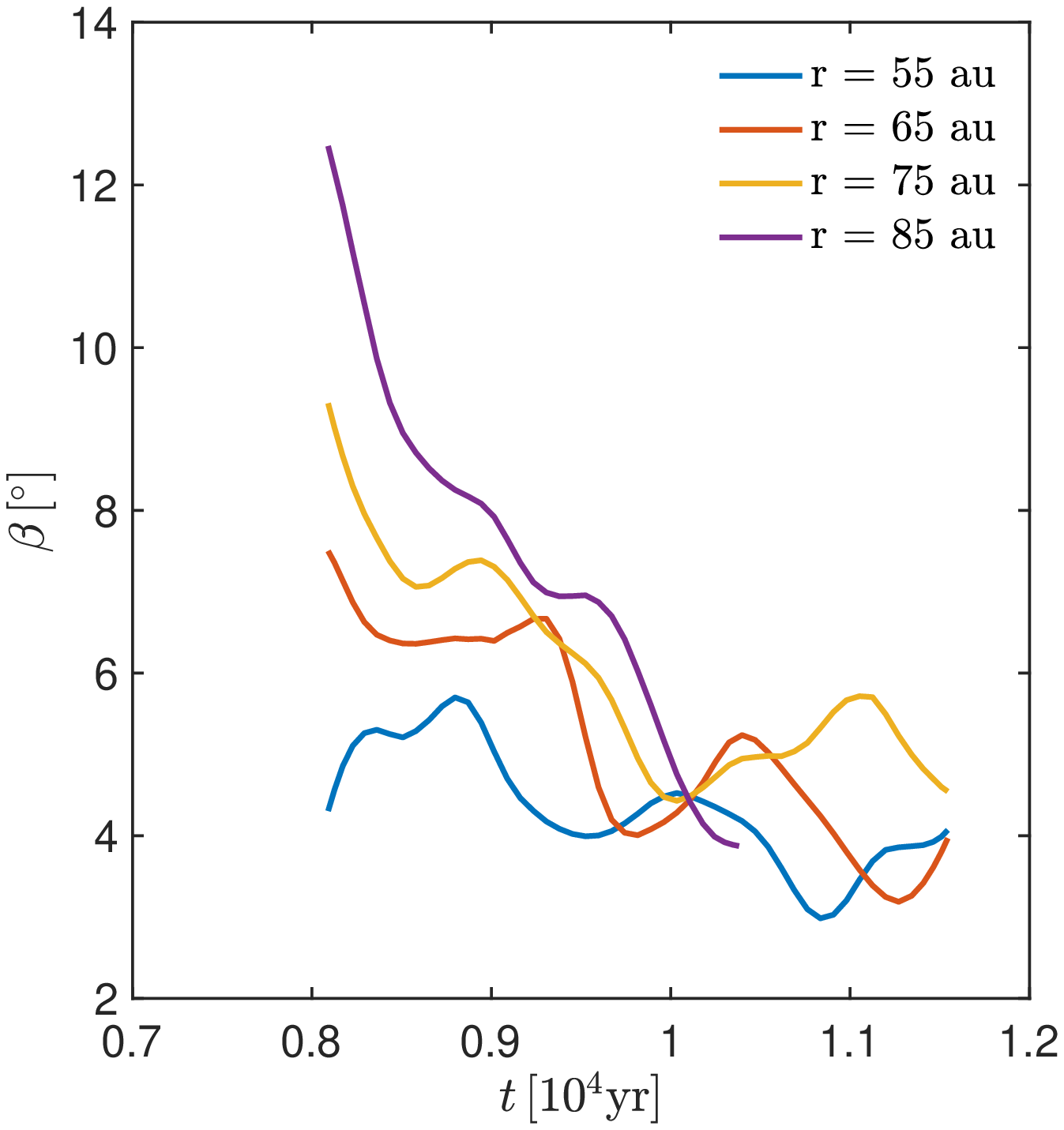}
\centering
\caption{The pitch angle, $\beta$, of arm 1 as a function of time in years for four radii in the disc, $55\, \rm au$ (blue), $65\, \rm au$ (red), $75\, \rm au$ (yellow) and $85\, \rm au$ (purple). The data is smoothed using a Gaussian filter over a ten-element sliding window. The radial locations match the colored symbols from Fig.~\ref{fig::pitch_spirals}.  The pitch angle decreases as the spiral winds up.  The analysis of the pitch angle at $r = 85\, \rm au$ stops prematurely due to the spiral arms winding up.  }
\label{fig::pitch_angle}
\end{figure}

 For the remainder of this subsection, we analyse only the two-armed spiral excited from the $100\, \rm M_J$ perturber since the arms are  the most conspicuous. Besides the pattern speed, we also calculate the pitch angle of the spirals. For simplicity, we only analyze the pitch angle for arm 1. The pitch angle, $\beta$, is defined as in \cite{Binney2008} with
\begin{equation}
    \tan \beta = \left| \frac{{\rm d}r}{r{\rm d}\theta} \right|\,,
    \label{eq::pitch_angle}
\end{equation}
where $r$ is the radial distance, and $\theta$ is the azimuthal position. The ${\rm d}r/{\rm d}\theta$ term is calculated from our simulations  by tracing the spiral arms in a ($r,\theta$) grid, using a similar method as in Fig.~\ref{fig::space_time}. To visualize the evolution of the pitch angle, we show the azimuthal surface density as a function of logarithmic radius at five different times in Fig.~\ref{fig::pitch_spirals}.  The top panel shows the time at periastron passage of the $100\, \rm M_J$ perturber (M100). We mark the peak surface density along arm 1 at the four chosen radii, $55\, \rm au$, $65\, \rm au$, $75\, \rm au$, and $85\, \rm au$. 
%If no spirals are present (top-left panel), we mark the azimuthal locations of the four radii at $\theta = +\pi $ for visualisation purposes only. 
The calculation of the pitch angles only begin when the spirals are excited. The perturber excites a two-armed spiral structure soon after the closest approach, shown in the second panel ($t = 8001\,\rm yrs$). As time increases, the spiral arms wind up, decreasing the pitch angle. The lower panels show times at which the spiral structure is winding up.  Eventually, the spiral arms dissipate, as shown in the bottom panel. After calculating ${\rm d}r/{\rm d}\theta$, we calculate the pitch angle evolution using Eq.~(\ref{eq::pitch_angle}) at $r = 55\, \rm au$, $65\, \rm au$, $75\, \rm au$, and $85\, \rm au$ given in  Fig.~\ref{fig::pitch_angle}. A Gaussian filter is applied to smooth the data. As the radial distance in the disc increases, the larger the initial pitch angle. At $r=55\, \rm au$, the initial pitch angle is $\beta \sim 4^\circ$, while at $r=85\, \rm au$, $\beta \sim 12^\circ$. The pitch angle decreases in time as the spiral winds up, regardless of radius,  which is consistent with the higher pattern speed closer to the star found in Fig.~\ref{fig::pattern_speed}. 

 \begin{figure} 
\centering
\includegraphics[width=\columnwidth]{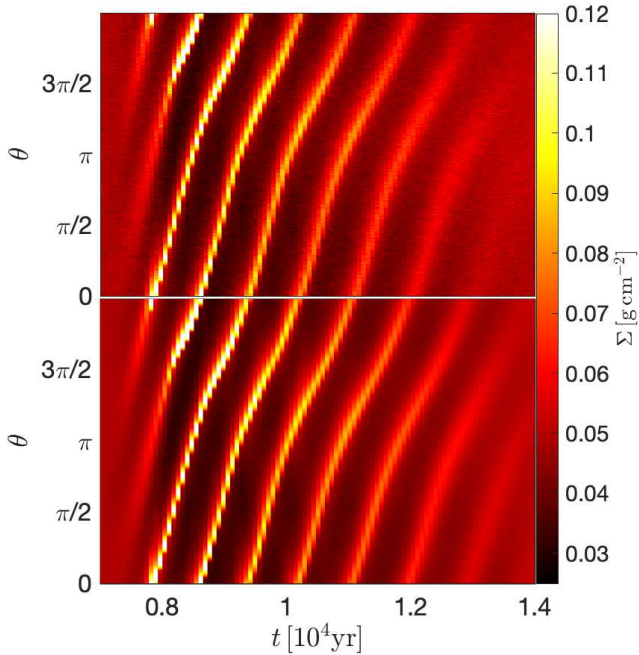}
\centering
\caption{ A resolution study showing the azimuthal disc surface density distribution at $r = 65\, \rm au$ for discs with $10^6$ particles (model M100, top sub-panel) and $10^7$ particles (M100h, bottom sub-panel).}
\label{fig::resolution1}
\end{figure}

\begin{figure} 
\centering
\includegraphics[width=.99\columnwidth]{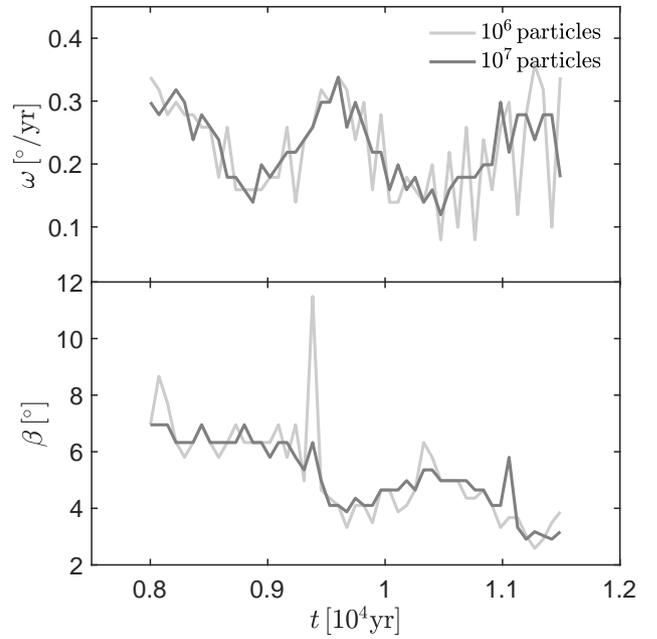}
\centering
\caption{A summary of our resolution study comparing models M100 ($10^6$ particles) and M100h ($10^7$ particles). We show the pattern speed ($\omega$) and pitch angle ($\beta$) for arm 1 at $65\, \rm au$ as a function of time for each simulation.}
\label{fig::resolution2}
\end{figure}

\begin{figure*} 
\includegraphics[width=2\columnwidth]{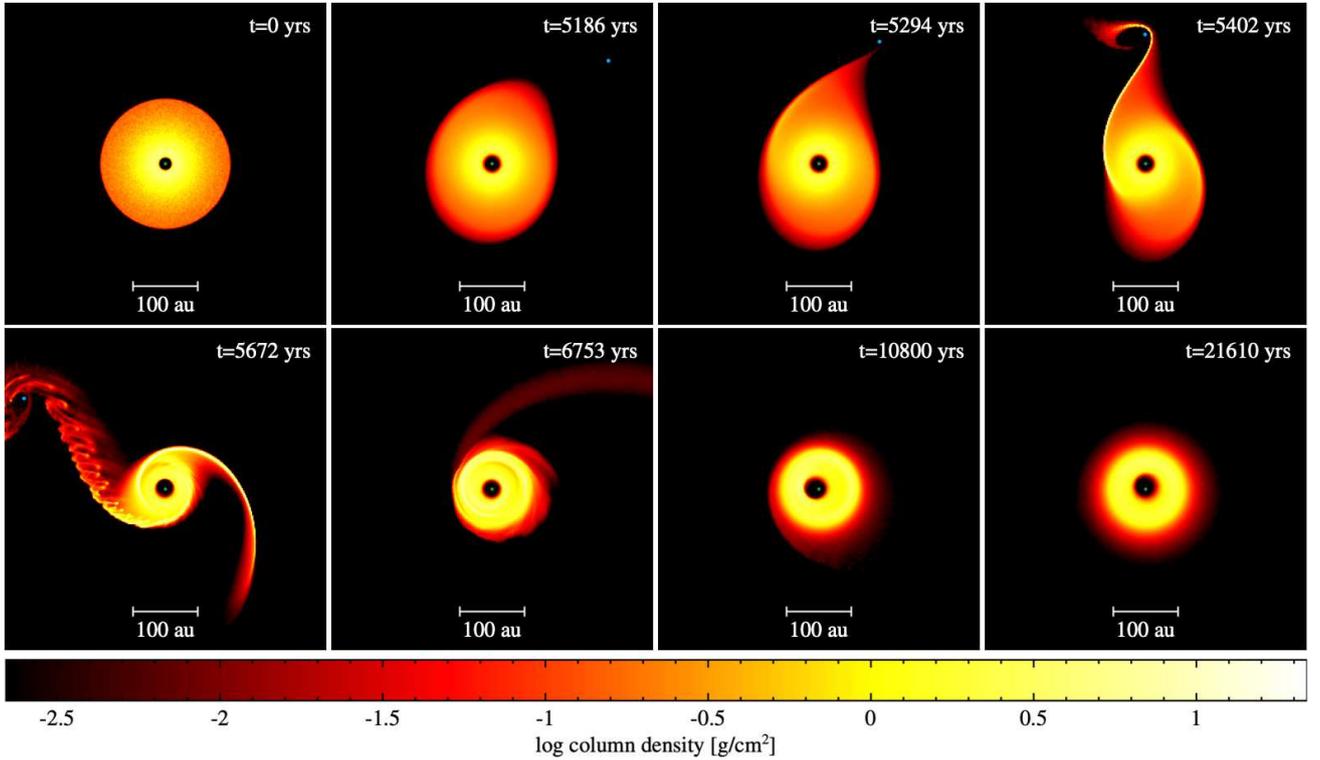}\centering
\caption{ Evolution of the disc structure under the influence of a stellar-mass perturber (M1000). We show the disc surface density at eight different times, with the top left panel being $t=0\, \rm yr$, and the top right panel being the time of closest approach of the perturber (blue dot shown in the north). The yellow is about two orders of magnitude denser than the red. The spiral arms quickly dissipates, leading behind a long-lasting eccentric cavity, shown in the bottom-right panels. 
%\nc{I am a bit puzzled by the wiggle in the spiral arm at 5672 yr, is this some sort of instability that develops along the arm? I've never seen something like that in my simulations, for similar orbital parameters and even with higher mass ratios (e.g. 2 and 5, see Fig. D1 in Cuello+2019.}
}
\label{fig::splash_stellarmass}
\end{figure*}

\begin{figure*} 
\centering
\includegraphics[width=0.98\columnwidth]{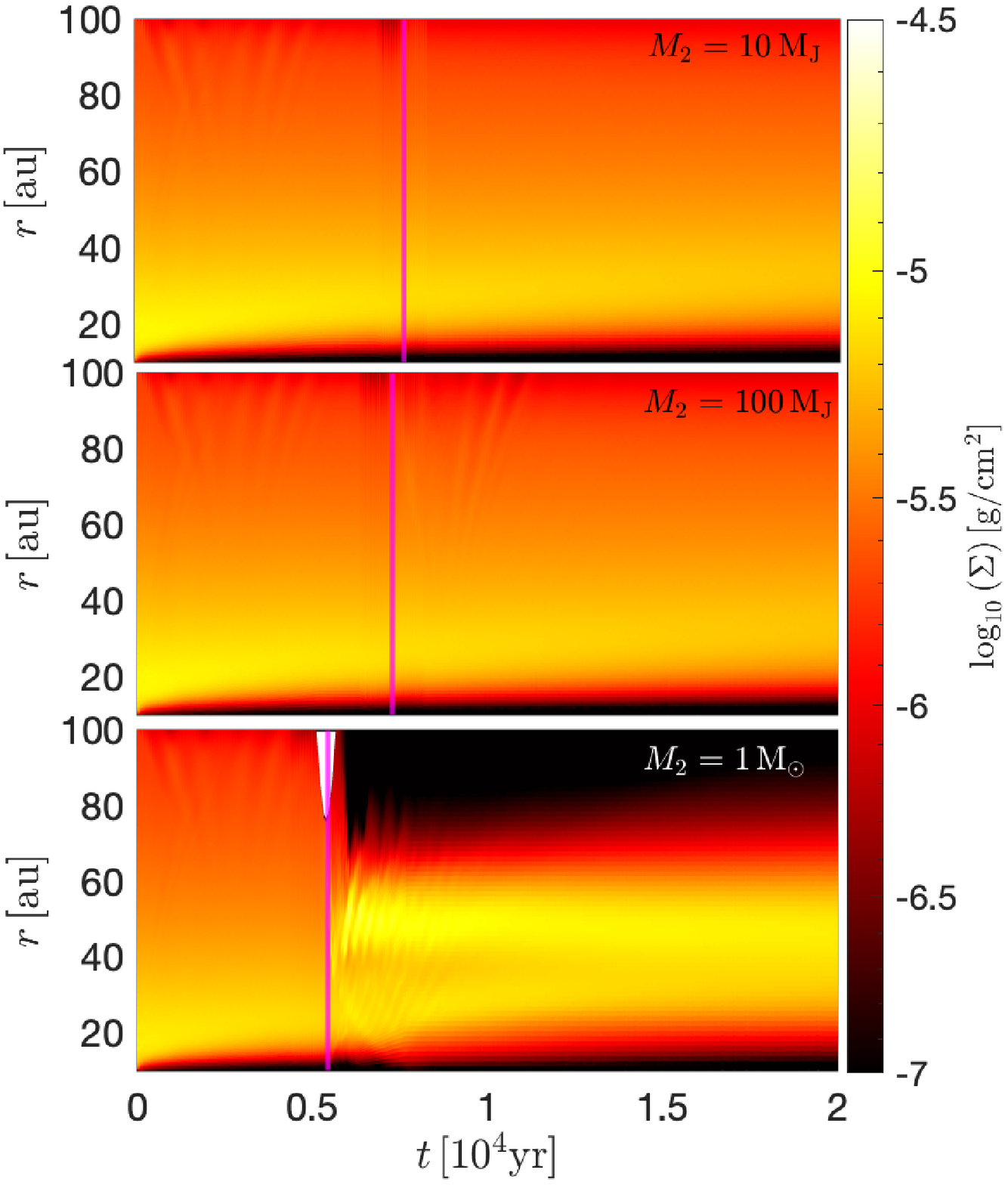}
\includegraphics[width=0.98\columnwidth]{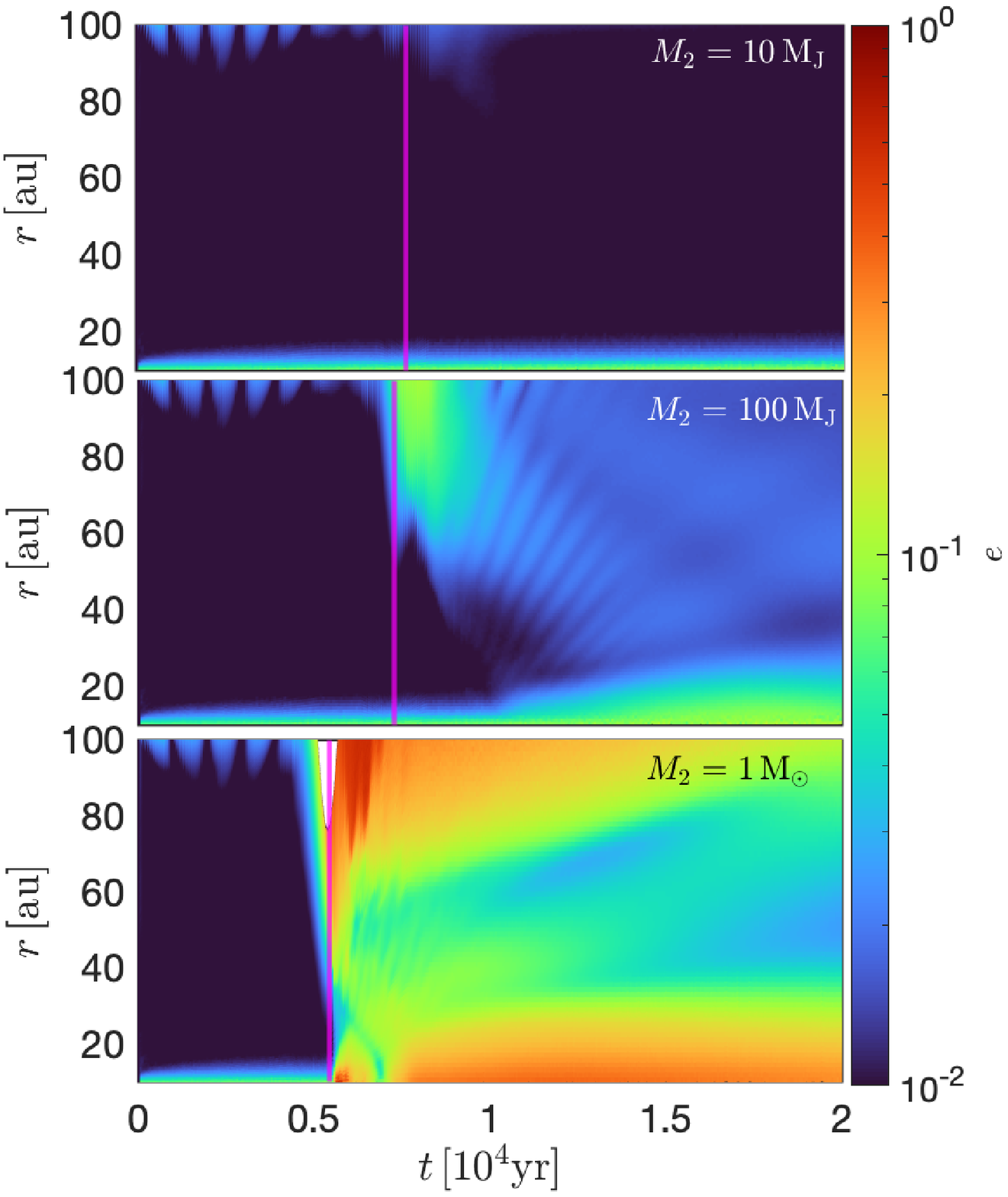}
\centering
\caption{Left panel: the disc surface density evolution for runs M10 (top sub-panel), M100 (middle sub-panel), and M1000 (bottom sub-panel), which corresponds to perturber masses of $10\, \rm M_J$, $100\, \rm M_J$, and $1\, \rm M_{\odot}$, respectively. The color map represents the logarithmic surface density, $\log_{10} \Sigma$, where $\Sigma$ is in g\,cm$^{-2}$. The vertical magenta line denotes the time of closest approach of the perturber in each simulation. Right panel: same as the left panel but for the disc eccentricity evolution, with the color map represents the disc eccentricity, $e$. A stellar-mass perturber on an unbound orbit causes heightened eccentricity growth within the protoplanetary disc. 
}
\label{fig::ecc_sigma}
\end{figure*}

\subsection{Resolution study} \label{SS:res}
 We present a resolution study for the simulation with a $100\, \rm M_J$ perturber. In smoothed particle hydrodynamical simulations, we make use of an artificial viscosity to model an expected \cite{Shakura1973} viscosity coefficient. However, the number of Lagrangian particles impacts how close the artificial viscosity is to the actual value. Here, we compare two simulations with $10^6$ particles (M100, $\alpha_{\rm AV} = 0.1260$) and $10^7$ particles (M100h, $\alpha_{\rm AV} = 0.2713$). Figure~\ref{fig::resolution1} shows the evolution of the disc surface density at 65\,au for the two simulations. The top sub-panel is the lower resolved disc, while the bottom sub-panel is the more highly resolved disc. The spirals in the higher resolution are  more conspicuous.  The lifetime of the spiral arms at higher resolution is longer, but only persists an extra few hundred years. Considering the total lifetime of the spirals for the lower resolution is $\sim 5000\, \rm yr$, the higher resolution spirals' lifetime is increased by less than $10$ per cent.  We expect this factor in spiral longevity to be greater for lower viscosity. In this work, we select a target viscosity of $\alpha_{\rm SS} \sim 0.005$, while observationally, protoplanetary discs can have viscosities in the range $10^{-3} - 10^{-5}$ \cite[e.g.,][]{Flaherty2015,Pinte2016,Flaherty2020,Villenave2022}. Protoplanetary discs can have higher viscosities depending on the disc size \cite[e.g.,][]{Hartmann1998,Ansdell2018}. Therefore, observed protoplanetary discs perturbed by flybys may have spiral arms that last longer than our numerical simulations presented in this work. 
%\nc{Based on your results, the effect is rather mild I would say. Spirals at high resolution can survive for an extra few hundred years. Considering the total lifetime of the spirals (5000 yr approx) we are underestimating their longevity by a factor of less than 10\%. Is this correct? If so, I would recommend being more quantitative.}

To further analyze simulations with a different initial number of particles, we compare the pattern speed and pitch angle for arm~1 between M100 and M100h in Fig.~\ref{fig::resolution2}. The more resolved simulation shows that the spiral arm surviving for a more extended period of time than the lower resolution spiral. For both the pattern speed and pitch angle, the lower resolution spiral mirrors the higher resolution spiral initially. However, deviations appear near the end of the low-resolution spiral's lifetime. This is due to the lowly resolved spiral beginning to dissipate. Despite that, the differences between the two simulations are minimal, and modeling the simulations with $10^6$ particles appears sufficient for capturing the dynamics at play.

\subsection{Coplanar equal mass  flyby}

Up to this point, we have only considered sub-stellar flyby perturbers. We now explore a stellar-mass encounter on an unbound parabolic orbit (M1000 from Table~\ref{table::setup}). Figure~\ref{fig::splash_stellarmass} shows snapshots of the disc structure during the stellar-mass flyby.  The top-left panel shows the initial disc structure at $t = 0\, \rm yr$. The top row panels slowly progress in time as the perturber reaches periastron in the top-right panel. At periastron, there is a tidal tail of material that extends from the primary disc to the perturber.  Shortly after the periastron passage (bottom-left panel), the perturber excites two prominent spiral arms with large pitch angles.  The arm extending towards the perturber shows substructures. We are uncertain whether these substructures are numerical or physical, i.e., an instability. We ran additional simulations with increased viscosity and particle resolution but the substructures are still present. The arms do not survive for long; however, the  inner cavity becomes elongated, which is consistent with the disc being eccentric. The eccentric disc structure survives for the duration of the simulation shown in the bottom-right panel, $\sim$15,000\,yr after periastron passage. The radial separation of the flyby at the end of the simulation is $\sim 4000\, \rm au$, meaning the disc is maintaining the eccentric structure even though the perturber is not close to perturb the disc.  This eccentricity growth induced in protoplanetary disc by flyby encounters is consistent with Fig.10 from \cite{Cuello2019}.
%\nc{FYI: we looked a bit this in our Flyby I paper. See for instance Fig. 10 where we report e vs r for different times: 540 yr and 2700 yr after pericentre. Your results are consistent with these trends. WONDERFUL FIGURE 11 BY THE WAY! I really like the temporal dimension of these plots. Well done!}
 We show further analysis of the short-lived spiral arms in the appendix.

We further investigate the eccentricity growth induced by a flyby event by comparing the disc surface density and eccentricity profiles for the $10\, \rm M_J$, $100\, \rm M_J$, and $1\, \rm M_\odot$ mass perturbers.  The left panel of Fig.~\ref{fig::ecc_sigma} shows the  azimuthally-averaged disc surface density as a function of time, with the color map denoting the surface density value. The top, middle, and bottom sub-panels are for a $10\, \rm M_J$, $100\, \rm M_J$, and $1\, \rm M_\odot$ mass perturber, respectively. The time of periastron passage of the encounter is shown by the vertical magenta line in each sub-panel.  The surface density profiles for the sub-stellar mass encounters are similar, with  little truncation of the outer radius of the disc. The bottom sub-panel shows the disc surface density for a stellar flyby. After the closest approach, the primary disc is strongly truncated by the perturber. The outer disc edge is truncated to $\sim 60\,\rm au$,  which is consistent with the tidal truncation radius of an equal-mass circular binary being one-third of the binary separation \citep{Artymowicz1994}. The right panel of Fig.~\ref{fig::ecc_sigma} shows the  azimuthally-averaged disc eccentricity as function of time, with the color map denoting the eccentricity value.  The eccentricity is determined by calculating the magnitude of the eccentricity vector for each radial bin. The stellar-mass perturber excites large eccentricity growth in the disc compared to the sub-stellar mass perturbers,  especially in the inner regions of the disc. The more massive perturber imparts more energy and angular momentum to the wave that are launched at Lindbald resonances. The disc then undergoes changes in angular momentum where the excited waves dampen, resulting in the disc eccentricity growth \cite[e.g.,][]{Lubow1991}. The eccentricity in the disc dampens over time due to pressure forces that recircularize the disc \citep{Pfalzner2003}.
The accretion radius of the sink has a minor effect on the eccentricity growth in the outer regions of the disc \cite[see appendix C in][]{Cuello2019}.

\begin{figure}
\centering
\includegraphics[width=\columnwidth]{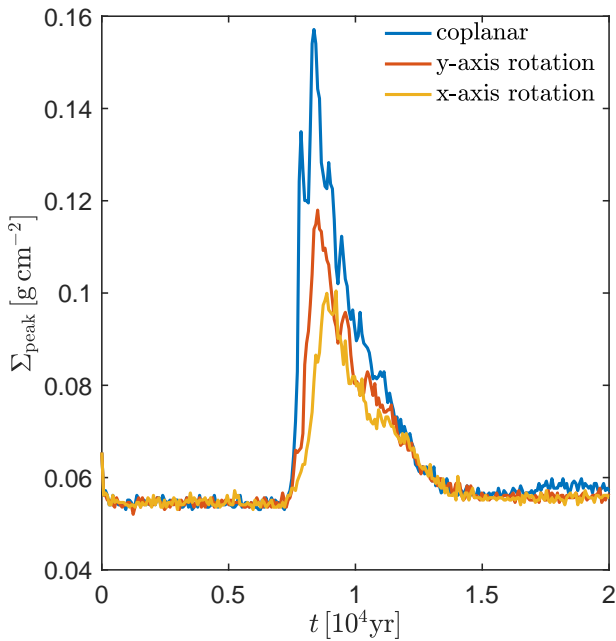}
\centering
\caption{The peak disc surface density, $\Sigma_{\rm peak}$,  as a function of time. The peak density is found by taking the maximum surface density over the disc azimuthal angle ($0 < \theta < 2\pi$) at $r = 65\, \rm au$. We consider a perturber mass of $100\, \rm M_J$ that is coplanar (blue) , $45^\circ$-inclined with respect to the $y$-axis (red), and $45^\circ$-inclined with respect to the $x$-axis (yellow). 
}
\label{fig::sigma_peak_inc}
\end{figure}

\begin{figure*} 
\centering
\includegraphics[width=2\columnwidth]{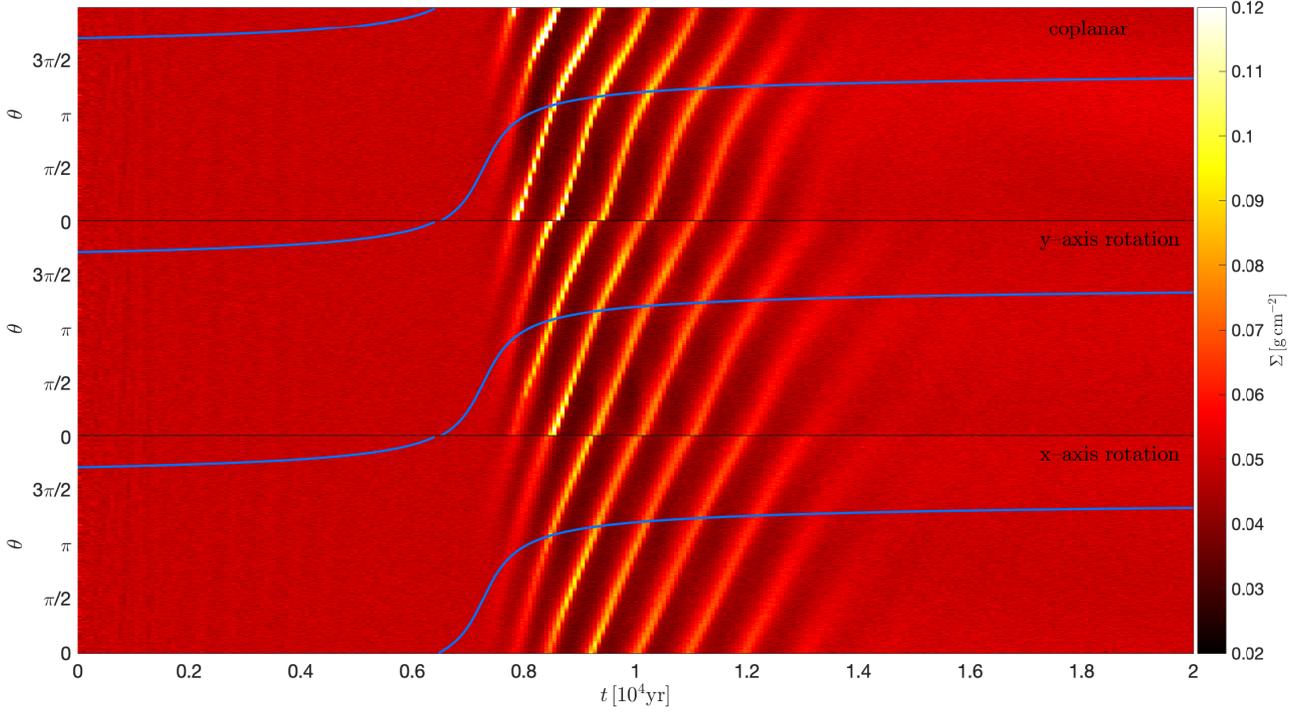}
\centering
\caption{An azimuthal cut of the disc surface density as a function of time in years for the coplanar (top panel, model M100), $y$-axis rotation (middle panel, M100iy), and $x$-axis rotation (bottom panel, M100ix) flyby simulations . The color map denotes the disc surface density. We analyze the surface density, $\Sigma$, at $r = 65\, \rm au$.  The azimuthal angle for the flyby orbit is given by the blue curve. The time of closest approach occurs at $\theta = +\pi/2$. The inclined models show similar spiral structure compared to the coplanar model.}
\label{fig::space_time_inc}
\end{figure*}

\begin{figure}
\centering
\includegraphics[width=0.99\columnwidth]{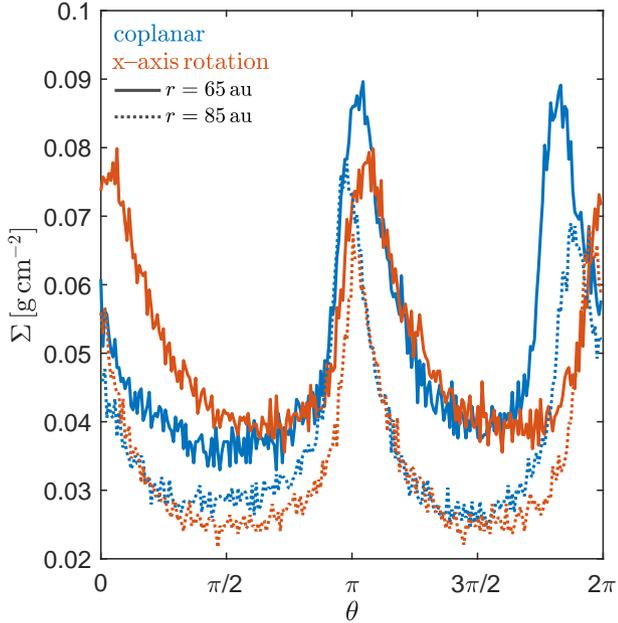}
\centering
\caption{Same as Fig.~\ref{fig::azimuthal_cut} but comparing the coplanar (red) and $x$--axis rotated (blue) flybys.}
\label{fig::azimuthal_cut_rot}
\end{figure}

\begin{figure}
\centering
\includegraphics[width=\columnwidth]{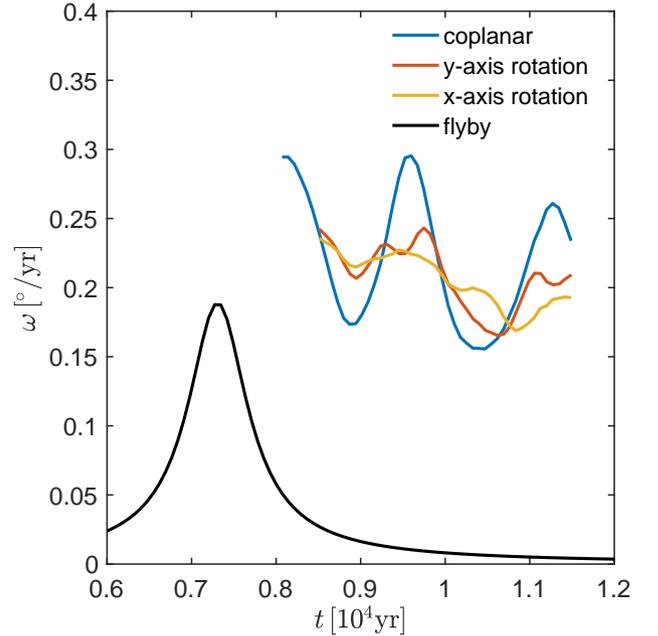}
\centering
\caption{The pattern speed, $\omega$, of arm 1 at $r = 65\, \rm au$ for the coplanar (M100, blue), y-axis rotated (M100iy, red), and  x-axis rotated (M100ix, yellow) simulations.  We smooth the data using a Gaussian filter over a ten-element sliding window. The angular velocity of the $100\, \rm M_J$ perturber is shown by the black curve. The peak corresponds to the time of closest approach of the flyby.}
\label{fig::pattern_speed_inc}
\end{figure}

\begin{figure}
\centering
\includegraphics[width=\columnwidth]{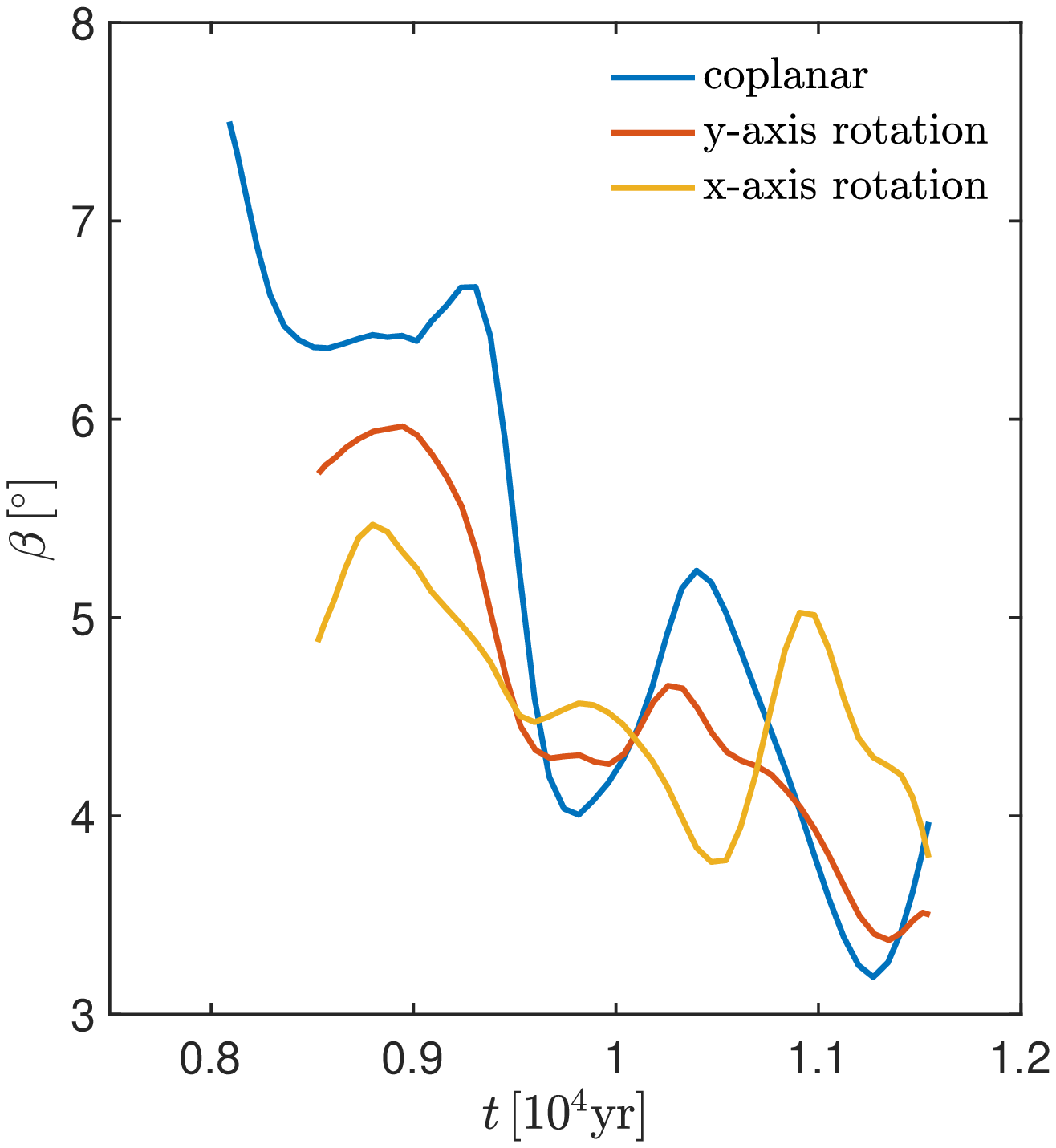}
\centering
\caption{The pitch angle, $\beta$, of arm 1 as a function of time in years in the disc at $65\, \rm au$ for the coplanar (M100, blue), y-axis rotated (M100iy, red), and  x-axis rotated (M100iy, yellow) simulations. The data is smoothed using a Gaussian filter over a ten-element sliding window.}
\label{fig::pitch_angle_inc}
\end{figure}

\subsection{Inclined flyby}
 We now investigate spiral arms excited by a $100\, \rm M_J$ flyby on an inclined trajectory with an initial misalignment of $45^\circ$. We consider two different orbital orientations, one rotated with respect to the $y$-axis and the other with respect to the $x$-axis. Figure~\ref{fig::sigma_peak_inc} shows the peak surface density, $\Sigma_{\rm peak}$,  at a radius $r = 65\, \rm au$ as a function of time for a coplanar, $y$-axis rotated, and $x$-axis rotated flyby. The maximum point of the curves corresponds to the strength of the spiral arms that are produced. A coplanar perturber excites a larger peak surface density. The $y$-axis rotated flyby excites the second highest peak surface density, followed by the $x$-axis rotated flyby. In Fig.~\ref{fig::azimuthal_cut_rot}, we show the spiral arm contrast versus azimuthal angle for the coplanar perturber (blue lines) and inclined perturber with a $x$--axis rotation (red lines). The solid lines denote the spiral arm at radius $r = 65\, \rm au$, and the dotted lines at $r = 85\, \rm au$. The two peaks in the surface density represent the individual arms. The density contrast between the arms is identical at each radius.

 Next, we compare the spiral structure that is excited within the disc between the coplanar and inclined models. Like the coplanar model, we select a time range to conduct the analyse to reduce the noise. For the rotation models, we analyse the spiral arms from $t = 8500\, \rm yr$ to $t = 11500\, \rm yr$.  Figure~\ref{fig::space_time_inc} shows an azimuthal cut of the disc surface density as a function of time. The top panel shows the coplanar flyby, the middle panel shows the $y$-axis rotation, and the bottom panel shows the $x$-axis rotation. Soon after periastron passage of the perturber (at $\theta = +\pi/2$), two spiral arms are excited within the disc for each model.  The spiral structure excited by the inclined cases are similar to the spirals excited by a coplanar flyby  of the same mass. For both inclined models,  arm 1 is first excited at a time of $\sim 7000\, \rm yr$, and lasts until $\sim 15000\, \rm yr$. The spiral arms generated from the coplanar companion are more prominent -- leading to a higher surface density concentration within the arms. Spirals excited by the $x$--axis rotated flyby are less prominent than spirals excited by the $y$--axis rotated flyby, as shown back in Fig.~\ref{fig::sigma_peak_inc}. 

 We analyse both the pattern speed and pitch angle of arm 1 for the inclined models and then compare to the coplanar case. The pattern speed and pitch angle calculations are performed  using the same techniques presented earlier. Figure~\ref{fig::pattern_speed_inc} shows the pattern speed of arm 1 at $65\, \rm au$ for each model, along with the angular velocity of the perturber. The initial pattern speed of arm 1 from each model are slightly higher than the perturber's angular velocity during periastron passage, but decreases over time. The pattern speed in the inclined cases are similar to the coplanar model but with less oscillations. 
 Figure~\ref{fig::pitch_angle_inc} shows the pitch angle of arm 1 a $65\, \rm au$ for the coplanar and inclined models. Similar to the coplanar case, the pitch angle for the inclined models decrease in time as the spiral arms wind up. As the flyby orbit is rotated about $y$-axis, the initial pitch angle as a function of radius is lower than the coplanar flyby simulation. The $x$-axis rotation model has a lower initial pitch angle than the $y$-axis rotation model. Overall, the inclined scenario shows similar spiral structures as the coplanar model.

\section{Discussion}
\label{sec::Discussion}

The interaction between the flyby and a protoplanetary disc can heat up the disc through compression and shocks \citep{Lodato2007a}. In our {\sc phantom} simulations, we do not take this heating into account. Protoplanetary discs with excited spirals can also be heated by absorbing stellar radiation, which produces a positive vertical temperature gradient \citep{Chiang1997}. The temperature gradient causes a thermal stratification in the disc, which can alter the morphology of the spiral wake \citep{Lee2015}. \cite{Juhasz2018} found that the pitch angle of the spirals produced by an embedded planet in a thermally stratified disc decreases near the disc mid-plane, and then increases towards the disc surface. These results can be applied to a stellar companion perturbing a disc, where the spiral pitch angles excited by a flyby event will be affected by a thermally stratified disc.  Moreover, previous hydrodynamics simulations of flyby-disc interactions used a radial surface density profile of $p=1$ to match observed disc profiles \cite[e.g.,][]{Cuello2019,Cuello2020}, which loads more material in the outer disc regions compared to $p=1.5$ used in this work. For $p=1$, we expect brighter spirals compared to $p=1.5$. Given the low disc mass simulated in this work, the dynamical behaviour of the disc during the encounter does not sensitively depend on the initial surface density profile. 
%If spiral arms are excited by flybys, the relationship between the pitch angle and local temperature  could be used to reconstruct the vertical thermal stratification of the disc \citep{Juhasz2018}. \

 The density variation in the spiral arms between our hydrodynamical models with different mass perturbers is about a factor of two, which may be distinguishable by observations. \cite{Cuello2020} provided observational signatures of flyby encounters. From their results, spiral arms excited by flybys can be observed in the continuum, which means the encounter is ongoing or the perturber is a few thousand  au from the host star. Moreover, the spiral arm between the two stars is brighter since it is more illuminated. Inclined encounters will excite spiral arms that are not coplanar with the disc, resulting in unique kinematic signatures that can be resolved with a spectral resolution of the order of $1\, \rm km/s$ to accurately map the vertical layers of the disc \citep{Cuello2019}. In \cite{Dong2017}, they showed spiral contrasts in scattered light (see their Fig. 5) versus spiral contrast in surface density (see their Fig. 1). They found a factor of two contrast in surface density roughly translates into a factor of two in scattered light surface brightness, and is detectable after convolving the images to achieve a realistic angular resolution. Even if the density enhancement is a factor of two at the midplane, it can be a lot higher at the surface \citep{Speedie2022}. The synthetic observations of flyby-induced spirals may provide direct evidence that they can be visible, \cite[e.g., Fig. 4 in][]{Cuello2020}. Therefore, the density enhancement of a factor of two in the spiral arms between our different mass perturber encounters should be distinguishable by observations. 

There are several observational examples of spiral arms in protoplanetary discs with no confirmed companion, with one being Elias 2-27 \cite[e.g.,][]{Perez2016}. As shown in this work and from previous works, a way to produce spirals without a bound companion is a flyby encounter. Another way to produce spirals without any perturber is gravitationally unstable  protostellar discs \cite[e.g.,][]{Baehr2021a}. Early evolution of a star-disc system suggest the mass of the central star is comparable to the disc mass, forcing the system to be in the self-gravitating regime \citep{Lin1987,Lin1990}. Numerical simulations of self-gravitating discs with a Toomre parameter $\mathcal{Q} \sim 1.5-1.7$ produced non-axisymmetric perturbations that grow into spiral density waves \citep{Durisen2007}. In the context of Elias 2-27, the spiral arms observed in the system are thought to be caused by the gravitational instability \citep{Meru2017,Tomida2017}. However, \cite{Duncan2018} postulated that a massive stable disc undergoing a stellar companion encounter can produce spirals with pitch angles that increase with radius \cite[e.g.,][]{Veronesi2021}.  Measuring the pattern speed of the spiral arms in Elias 2-27 offers a more direct assessment of the origin of the spirals. If the pattern speed of the spirals decreasing in time, Elias 2-27 may have undergone a flyby in the past. A single value for each spiral's pitch angle was measured, $\beta \sim 15^\circ$, by \cite{Huang2018} without a radial dependence. Currently, the measured values do not favour a specific scenario.  \cite{Shuai2022} used the  Gaia Data Release 3 (DR3) to determine if flyby encounters occurred for 20 systems with spiral arms. For their selected systems, their analysis suggested that an unbound encounter was not the cause for isolated spiral arms in scattered light. We note that Gaia DR3 does not have radial velocity information for three-dimensional motion tracing and thus only can map two-dimensional on-sky locations of the systems. More observations are needed to ascertain the origin of the spirals.

 The spirals excited by a flyby perturber are different from the GI spirals and spirals excited by a bound perturber in several ways. The pitch angle of GI spirals are $\sim 10^{\circ}$ throughout the disc  \citep{Baehr2021a,Bethune2021}, The pitch angle of spirals excited by either a bound or unbound perturber decreases further away from the perturber \citep{Zhu2015b,Bae2018}. As shown in this work, the spirals from an unbound perturber would wind up and disappear after several orbits. The spirals from a circularly orbiting perturber have a steady shape, co-orbiting with the perturber. The spirals from an eccentric perturber would also evolve and change its shape with time but normally have multiple complex spiral arms \citep{Zhu2022}.

The hydrodynamical simulations conducted in this work only dealt with one specific disc size, which modeled a non-penetrating encounter. Increasing the disc size makes the model approach a grazing encounter. During a grazing encounter, spiral arms should still be produced within the primary disc, however, material may be captured by the perturber, and form a secondary disc \citep{Breslau2017,Cuello2019}. If the outer disc edge encloses the periastron of the flyby, we then have a penetrating encounter. During a penetrating encounter, the primary disc size should be truncated, and accretion outbursts may occur due to the flyby \citep{Vorobyov2020,Borchert2022,Borchert2022b}. As shown by our resolution study (Section~\ref{SS:res}), viscosity affects the lifetime of the spiral arms. Less viscous discs that encounter a flyby should maintain spiral features and disc substructures on a longer timescale than more viscous discs.

From our hydrodynamical simulations,  flyby encounters can significantly affect the structure of protoplanetary discs, which has implications for planet formation. Dust grains can become trapped within spiral arms due to local gas pressure maxima \citep{Haghighipour2003a,Haghighipour2003b,Baehr2021a}. Dust trapping could promote grain growth, and eventually planetesimal formation \citep{Baehr2022}. Dust particles with Stokes number $\sim 1$ are more prone to become trapped at the local pressure maxima in protoplanetary discs \citep{Rice2006,Paardekooper2006}. \cite{Thies2010} demonstrated that flyby encounters may trigger gravitational instability, which can lead to the formation of brown dwarfs and giant planets in the outer regions of the disc. Hence, massive companions or planets can be formed at great distances from their host stars.

 Perturbers with few tens of Jupiter masses are able to excite spiral arms and disc substructures, including eccentricity growth. Another key observational signature in a disc that has been recently perturbed by a flyby is the presence of disc eccentricity that damps over time (see Fig.~\ref{fig::ecc_sigma}). A significant disc eccentricity can leave an observable imprint in the CO emission line profiles, which appears as an asymmetry in the line profile \citep{Regaly2011}.  For example, \cite{Flaherty2015} observed a small asymmetry in the CO line profiles of the HD 163296 protoplanetary disc, which may be consistent with the disc being eccentric.  In the case of disc eccentricity excited by a flyby event, the observed asymmetry in the CO line profiles should slowly dissipate over time as the disc eccentricity dampens.  However, given that the eccentric disc around HD 163296 is on large scale, the timescale for observable eccentricity decrease should be longer than the detectability timescale of $\sim 10\, \rm yr$ (operating time for ALMA). 

   Based on the results of this work, the changes in the spiral structure occur over timescales of thousands of years, which are not observable in any practice sense. However, it may be possible to distinguish the origin of spiral arms at a single epoch by measuring the pitch angle of the spirals \cite[see Table 1 from][]{Yu2019} . A disc perturbed by a flyby should eventually have more tightly wound spirals than spiral arms excited by bound companions or GI. The pitch angle of spirals excited by both bound companions \citep[the inner arms;][]{Dong2015b}
  %several tens of degrees \cite[see Table 1 from][]{Yu2019} 
  and GI \citep{Baehr2021a,Bethune2021} can be $\gtrsim 10^{\circ}$. 
  %Therefore, instead of observing multiple epochs of spiral structure which is difficult, we may be able to estimate the timing of past flyby encounters by measuring how tightly wound the spirals are \cite[see Table 1 from][]{Yu2019} .

  Is it possible to disentangle the effects of a bound binary from a flyby event? Binary stars will excite eccentricity growth in circumbinary discs \cite[e.g.,][]{Dunhill2015,Ragusa2020,Smallwood2022,Siwek2022}.  Distinguishing these two mechanisms at a single epoch would be rather challenging. However, thanks to kinematic measurements of stellar masses and recent Gaia data, it is possible to better constrain the orbits of (supposedly) interacting stars. This is done for instance in \cite{Weber2023} (see their Fig. 11) for the system AS205N and AS205S. For that specific case, despite the degeneracy, they determine that the orbit is very likely parabolic/hyperbolic, not bound. In addition, for the flyby case, the eccentric disc eventually circularizes once the flyby leaves the system. Moreover, if the dynamical center of an eccentric disc (i.e., the foci of the disc) can be determined precisely, perhaps in the circumbinary disc cases one can see that the disc is not dynamically centered on the primary star (often the more visible star). The spirals excited by bound stellar companions \cite[e.g.,][]{Rosotti2020,Gonzalez2020,Poblete2020,Poblete2022} evolve differently to spiral excited by stellar flybys. For example, the pattern speed and  pitch angle of the spiral arms should remain rather constant in time with a bound companion. This is in direct contrast to flyby events, where the pattern speed and pitch angle evolve in time.

 Flybys -- although considered as violent and rather destructive events -- can generate favourable conditions for planet formation in the long run. After disc truncation, the surface density of the disc is enhanced locally in regions where the eccentricity remains low (see Fig.~\ref{fig::ecc_sigma}). In any case, the eccentricity excitation would dampen due to pressure forces in the disc that causes the disc to recircularise \citep{Pfalzner2003}. Once the eccentricity excitation dissipates, the disc becomes more compact and denser, which can be a favourable environment for planet formation. \cite{Cuello2019} found that removing gas and small dust grains at large disc radii effectively leads to higher values of dust-to-gas ratios in the inner parts of the disc. Speculatively, this could lead to the formation of self-induced dust traps \citep{Gonzalez2017,Vericel2021} or potentially trigger planetesimal formation \citep{JYM09,BS10,YJC17,LY21,SJL21}.

\section{Conclusions}
\label{sec::Conclusions}

We investigate the interaction of a protoplanetary disc with a flyby using three-dimensional SPH simulations. The unbound perturber excites a two-armed spiral structure in the disc. We model both sub-stellar and stellar flybys, as well as coplanar and inclined flyby trajectories. For the sub-stellar flybys, the more massive the perturber, the more prominent the spiral arms. The pattern speed of the spiral arms excited by coplanar and inclined flybys are close to the angular velocity of the perturber at periastron.  Overall, spiral arms last for a couple of dynamical timescales at the radius of the periastron before winding up and disappearing.  From our resolution study, we found that disc viscosity plays a role in the lifetime of flyby-included spiral arms. Discs with lower viscosity maintain the spiral arms for a longer period of time versus discs with higher viscosity.  The stellar mass flyby strongly perturbs the disc, forcing significant eccentricity growth, which dampens overtime. 

  From our hydrodynamical simulations of flyby interactions, the pattern speed and pitch angle of the flyby-induced spiral arms evolve in time. Both the pattern speed and pitch angle of spiral arms decrease over time as the spiral arms wind up. 
Therefore, the pitch angle could be a unique observation signature for flyby-induced spirals.  The pattern speed, pitch angle, and general morphology of the spirals excited by an inclined flyby are similar to a coplanar flyby. For typical flyby configurations, the temporal window to search for nearby perturbers is of the order of a few kyr (tens of kyr at most). Here, we postulate that the spirals' morphology and the disc eccentricity can be used to search for / chase potential unbound stars or planets around discs where a flyby is suspected. Future disc observations at high resolution and dedicated surveys will help to constrain the frequency of such stellar encounters in nearby star forming regions.

\section*{Acknowledgements}
 We thank the anonymous referee for helpful suggestions that positively impacted the work.  We acknowledge the use of SPLASH \citep{Price2007} for the rendering of the figures. JLS acknowledges funding from the ASIAA Distinguished Postdoctoral Fellowship. CCY and ZZ acknowledge the support by NASA via the Emerging Worlds program (grant number 80NSSC20K0347) and the Astrophysics Theory Program (grant number 80NSSC21K0141). 
CCY is also grateful for the support by NASA TCAN program (grant number 80NSSC21K0497).
RGM acknowledges support from NASA through grant 80NSSC21K0395.
This research was supported by the Munich Institute for Astro-, Particle and BioPhysics (MIAPbP) which is funded by the Deutsche Forschungsgemeinschaft (DFG, German Research Foundation) under Germany´s Excellence Strategy -- EXC-2094 -- 390783311.
This research was funded, in part, by ANR (Agence Nationale de la Recherche) of France under contract number ANR-22-ERCS-0002-01. This project has received funding from the European Research Council (ERC) under the European Union Horizon 2020 research and innovation program (grant agreement No. 101042275, project Stellar-MADE).

%%%%%%%%%%%%%%%%%%%%%%%%%%%%%%%%%%%%%%%%%%%%%%%%%%
\section*{Data Availability}
The data supporting the plots within this article are available on reasonable request to the corresponding author. A public version of the {\sc phantom} and {\sc splash} codes are available at \url{https://github.com/danieljprice/phantom} and \url{http://users.monash.edu.au/~dprice/splash/download.html}, respectively.



\bibliographystyle{mnras}
\bibliography{ref.bib} % if your bibtex file is called example.bib

%\begin{thebibliography}{99}
%\end{thebibliography}

%%%%%%%%%%%%%%%%%%%%%%%%%%%%%%%%%%%%%%%%%%%%%%%%%%

%%%%%%%%%%%%%%%%% APPENDICES %%%%%%%%%%%%%%%%%%%%%

\appendix

 \section{equal-mass flybys}

 \begin{figure*} 
\centering
\includegraphics[width=2\columnwidth]{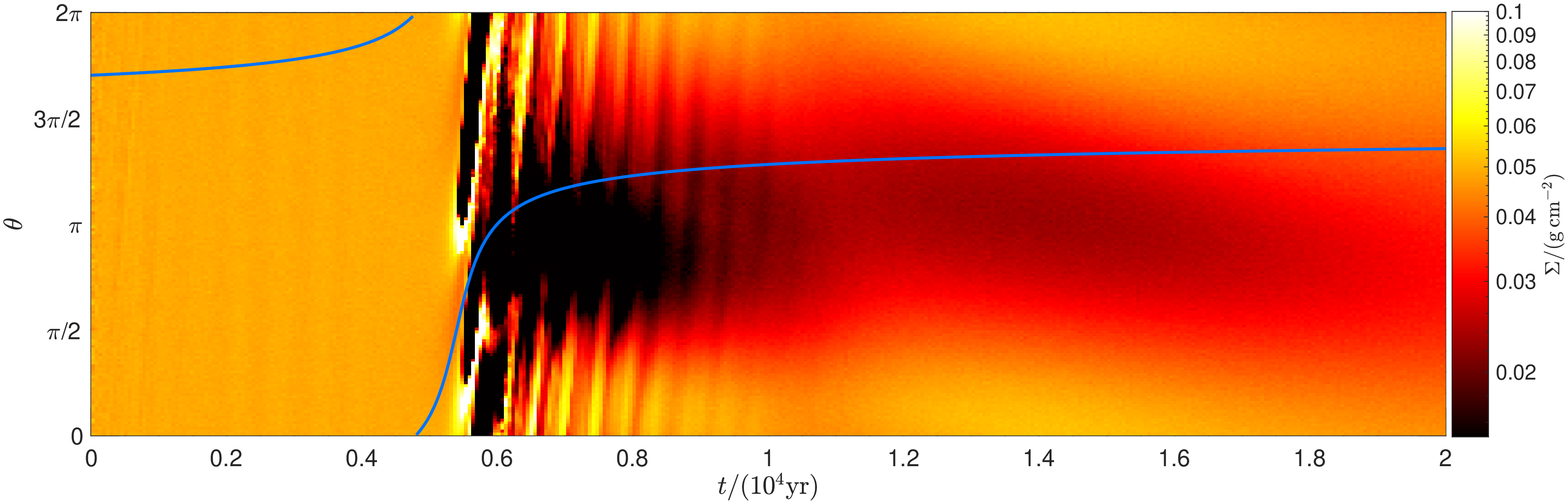}
\centering
 \caption{An azimuthal cut of the disc surface density as a function of time in years for a $1\, \rm M_{\odot}$ perturber (model M1000 from Table~\ref{table::setup}). The colour map denotes the disc surface density. We analyze the azimuthal angle $\theta$ at $r = 65\, \rm au$.  The azimuthal angle for the flyby orbit is given by the blue line. The time of closest approach occurs at $\theta = +\pi/2$.}
\label{fig::space_time_run6}
\end{figure*}

 We examine the azimuthal  variation of the surface density at a given radius for the equal-mass perturber simulation. Figure~\ref{fig::space_time_run6} shows the azimuthal cut of the disc surface density at a radius of $r = 65\, \rm au$. The angular position of the flyby is given by the solid blue curve. The time of periastron passage occurs at $\theta = +\pi/2$ in each simulation. Before the time of the closest approach of the perturber, the protoplanetary disc  appears quiescent, with no excited substructures. Soon after periastron passage, the perturber excites a two-armed spiral within the disc, seen by the  azimuthal striations in the surface density. Unlike the unequal mass simulations, the two-armed spiral is short-lived, only lasting until $\sim 6000\, \rm yr$.  Unlike the unequal-mass models, soon after the two-arm spirals are excited, the arm led by the perturber develops wiggles and then merges into the other arm.  The remaining one-arm structure is short-lived, lasting for about 6000 yrs.  In the mean time, the perturber excites eccentricity growth within the disc. 

\begin{figure} 
\centering
\includegraphics[width=1\columnwidth]{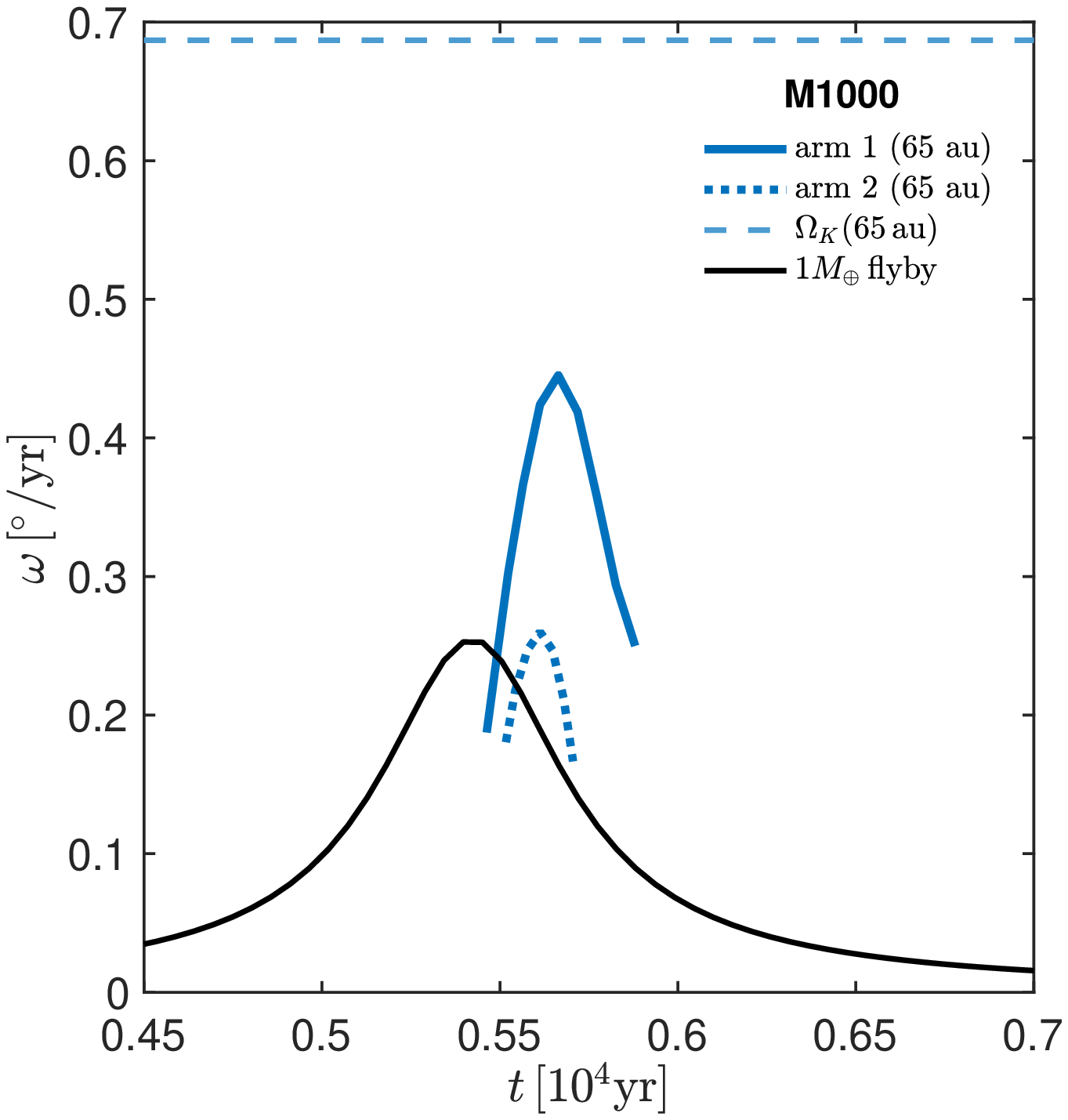}
\centering
\caption{The pattern speed, $\omega$, of the two-armed spiral at $r = 65\, \rm au$ (blue) for a $1\, \rm M_{\odot}$ perturber (M1000).  We smooth the data using a Gaussian filter over a ten-element sliding window. The pattern speed for arm 1 is given by the solid curves, and arm 2 by the dotted curves. The angular velocity of the perturber is shown by the black curve. The peak corresponds to the time of closest approach of the flyby. The Keplerian angular velocity, $\Omega_{\rm K}$, at $r = 65\, \rm au$ is shown by the horizontal blue dashed line.}
\label{fig::pattern_speed_run6}
\end{figure}

Figure~\ref{fig::pattern_speed_run6} shows the pattern speed of arm 1 (solid curve) and arm 2 (dotted curve) as a function of time  for the equal-mass perturber simulation. We calculate the pattern speed at $65\, \rm au$. We smooth the data using a Gaussian filter.  The angular velocity of the perturber is shown by the black curve and the Keplerian angular velocity at radii $r = 65\, \rm au$ is shown by the horizontal dashed line.  The initial pattern speeds of the arms are closely set by the angular velocity of the perturber at periastron. The pattern speeds of the arms also display an oscillation in the amplitude, similar to Fig.~\ref{fig::pattern_speed}.

\begin{figure} 
\centering
\includegraphics[width=1\columnwidth]{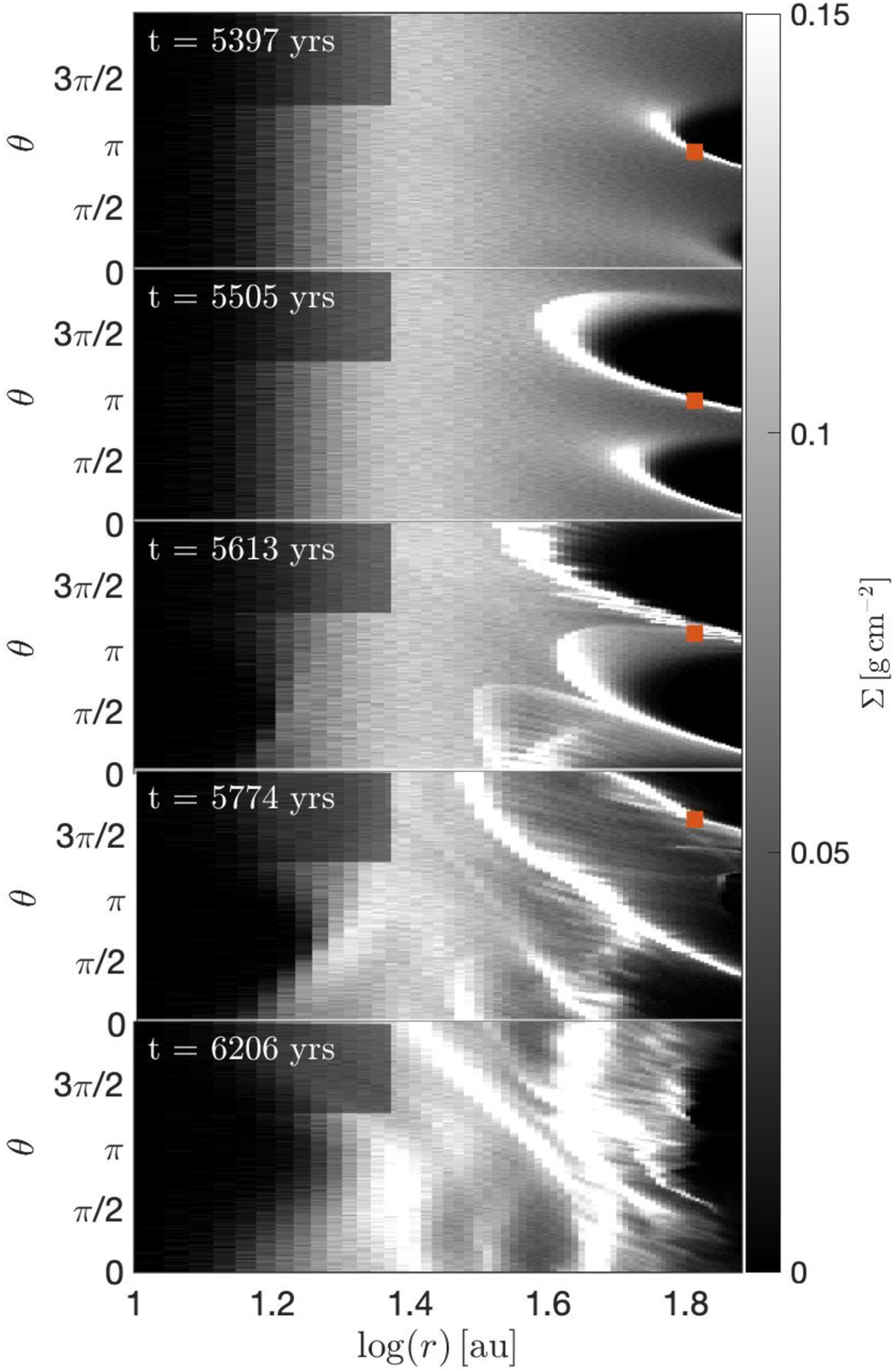}
\centering
\caption{Surface density in an ($r$,$\theta$) grid. The maps show the surface density in units of $\rm g\, cm^{-2}$. We show the surface density at  five different times, with the upper panel being the time when the perturber is at closest approach. We mark the peak surface density along arm 1 at $65\, \rm au$ (red square).}
\label{fig::r_theta_run6}
\end{figure}

\begin{figure} 
\centering
\includegraphics[width=1\columnwidth]{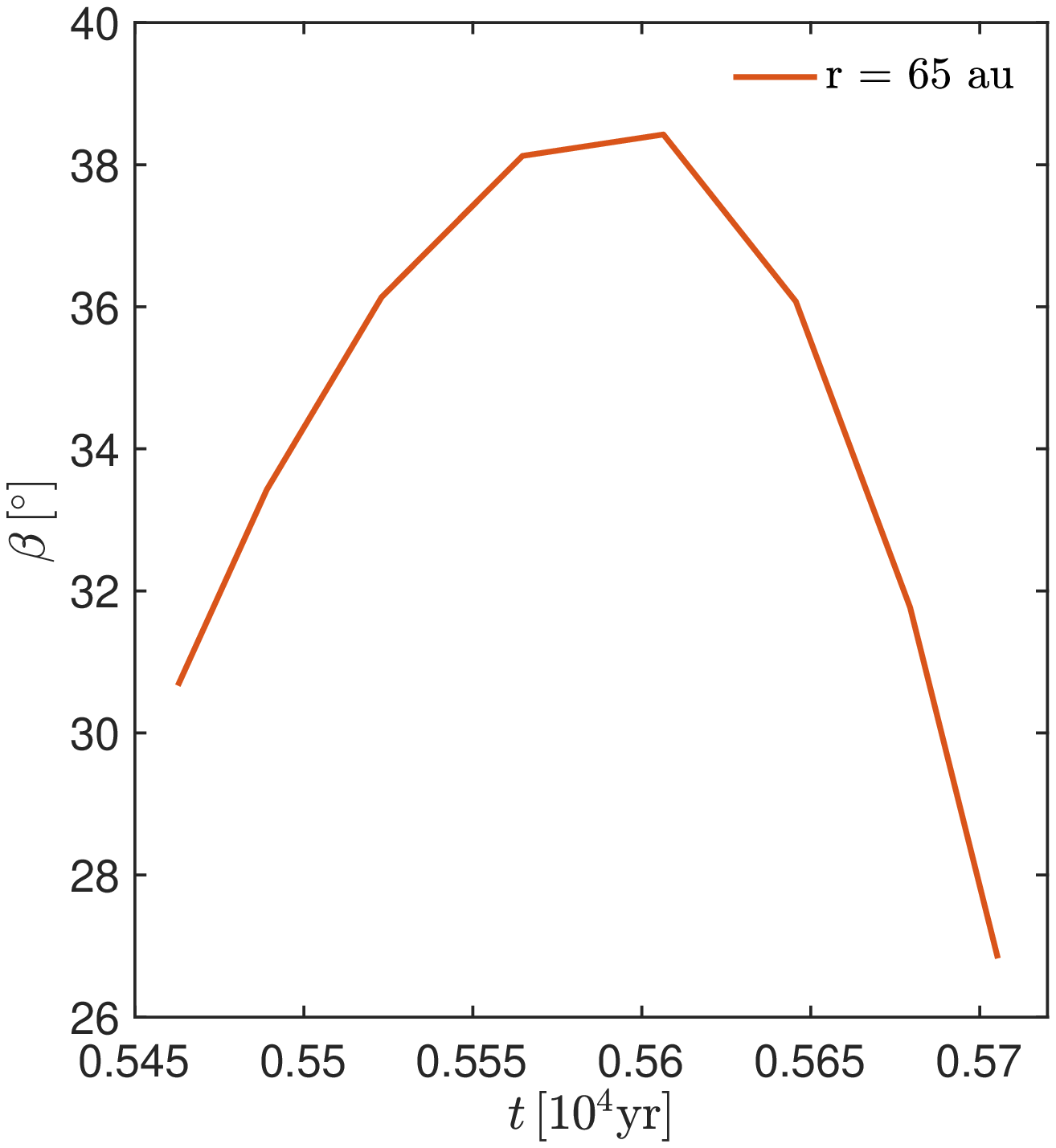}
\centering
\caption{The pitch angle, $\beta$, of arm 1 as a function of time in years at $r = 65\, \rm au$  for a $1\, \rm M_{\odot}$ perturber (M1000).  We smooth the data using a Gaussian filter over a ten-element sliding window. The radial location matches the colored symbol from Fig.~\ref{fig::r_theta_run6}.}
\label{fig::pitch_angle_run6}
\end{figure}

To visualize the evolution of the pitch angle, we show the azimuthal surface density as a function of logarithmic radius at five different times in Fig.~\ref{fig::r_theta_run6}.  The top panel shows the time at periastron passage of the $1\, \rm M_{\odot}$ perturber (M1000). We mark the peak surface density along arm 1 at the radii, $r = 65\, \rm au$. The calculation of the pitch angles only begins when the spirals are excited. The perturber excites a two-armed spiral structure, shown in the top panel. As time increases, the initial spiral arms are short-lived. The lower panels show times at which the spiral structure is winding up and also begin having a chaotic evolution. We calculate the pitch angle evolution using Eq.~(\ref{eq::pitch_angle}) at $r = 65\, \rm au$, given in  Fig.~\ref{fig::pitch_angle_run6}. We smooth the data using a Gaussian filter. The pitch angle first increases in time as the spiral arms are excited and then begins to decrease in time as the spiral winds up.

% If you want to present additional material which would interrupt the flow of the main paper,
% it can be placed in an Appendix which appears after the list of references.

%%%%%%%%%%%%%%%%%%%%%%%%%%%%%%%%%%%%%%%%%%%%%%%%%%

% Don't change these lines
\bsp	% typesetting comment
\label{lastpage}
\end{document}